\documentclass[trackchanges,twocolumn]{aastex702}

\usepackage{graphicx}	
\usepackage{amsmath}	
\usepackage{float}
\usepackage{multirow} 
\usepackage{cleveref}
\usepackage{enumitem}
\usepackage{array}
\usepackage{booktabs}
\usepackage{makecell}
\usepackage{placeins}
\usepackage{orcidlink}
\setlist[enumerate]{label=\arabic*.}

\defcitealias{2026ApJS..282....3S}{Sarrouh \& Asada et al.}

\begin{document}

\title{A Systematic Assessment of Total Stellar Mass Measurements with Spatially Resolved Photometry and Medium Bands at 1 $< z <$ 9: The Impact of Mass-to-Light Variations and Outshining}

\author[orcid=0009-0009-9848-3074]{Naadiyah Jagga}
\affiliation{Department of Physics and Astronomy, York University, 4700 Keele St. Toronto, Ontario, M3J 1P3, Canada}
\email[show]{jagga@yorku.ca}  

\author[orcid=0000-0002-9330-9108]{Adam Muzzin} 
\affiliation{Department of Physics and Astronomy, York University, 4700 Keele St. Toronto, Ontario, M3J 1P3, Canada}
\email{muzzin@yorku.ca}

\author[orcid=0000-0003-0780-9526]{Visal Sok}
\affiliation{Department of Astrophysical and Planetary Sciences, University of Colorado, 2000 Colorado Ave, Boulder, CO 80309, USA}
\email{visal.sok@colorado.edu} 

\author[orcid=0000-0002-3503-8899]{Vivian Yun Yan Tan}
\affiliation{Department of Physics and Astronomy, York University, 4700 Keele St. Toronto, Ontario, M3J 1P3, Canada}
\email{tanvivia@yorku.ca} 

\author[orcid=0000-0001-8830-2166]{Ghassan Sarrouh}
\affiliation{Department of Physics and Astronomy, York University, 4700 Keele St. Toronto, Ontario, M3J 1P3, Canada}
\email{gsarrouh@yorku.ca} 

\author[orcid=0000-0001-9298-3523]{Kartheik Iyer}
\affiliation{Columbia Astrophysics Laboratory, Columbia University, 550 West 120th Street, New York, NY 10027, USA}
\email{kgi2103@columbia.edu} 

\author[orcid=0000-0003-3983-5438]{Yoshihisa Asada}
\affiliation{Dunlap Institute for Astronomy and Astrophysics, 50 St. George Street, Toronto, Ontario, M5S 3H4, Canada}
\email{yoshi.asada@utoronto.ca} 

\author[orcid=0000-0002-4201-7367]{Chris J. Willott}
\affiliation{National Research Council of Canada, Herzberg Astronomy \& Astrophysics Research Centre, 5071 West Saanich Road, Victoria, BC, V9E 2E7, Canada}
\email{chris.willott@nrc.ca} 

\author[orcid=0000-0001-5984-0395]{Maru\v{s}a Brada\v{c}}
\affiliation{Faculty of Mathematics and Physics, Jadranska ulica 19, SI-1000 Ljubljana, Slovenia}
\affiliation{Department of Physics and Astronomy, University of California Davis, 1 Shields Avenue, Davis, CA 95616, USA}
\email{marusa.bradac@fmf.uni-lj.si} 

\author[orcid=0000-0003-3243-9969]{Nicholas S. Martis}
\affiliation{Faculty of Mathematics and Physics, Jadranska ulica 19, SI-1000 Ljubljana, Slovenia}
\email{nicholas.martis@fmf.uni-lj.si} 

\author{Ga\"el Noirot}
\affiliation{Space Telescope Science Institute, 3700 San Martin Drive, Baltimore, Maryland 21218, USA}
\email{gnoirot@stsci.edu} 

\author[orcid=0009-0009-4388-898X]{Gregor Rihtar\v{s}i\v{c}}
\affiliation{Faculty of Mathematics and Physics, Jadranska ulica 19, SI-1000 Ljubljana, Slovenia}
\email{gregor.rihtarsic@fmf.uni-lj.si} 

\author[orcid=0000-0002-7712-7857]{Marcin Sawicki}
\affiliation{Department of Astronomy and Physics and Institute for Computational Astrophysics, Saint Mary's University, 923 Robie Street, Halifax, Nova Scotia B3H 3C3, Canada}
\email{marcin.sawicki@smu.ca} 

\author[orcid=0009-0000-8716-7695]{Sunna Withers}
\affiliation{Department of Physics and Astronomy, York University, 4700 Keele St. Toronto, Ontario, M3J 1P3, Canada}
\email{sunnaw@yorku.ca} 

\author[orcid=0000-0002-0243-6575]{Jacqueline Antwi-Danso}
\affiliation{Dunlap Institute for Astronomy and Astrophysics, 50 St. George Street, Toronto, Ontario, M5S 3H4, Canada}
\affiliation{Department of Astronomy, University of Massachusetts Amherst, 710 North Pleasant Street, Amherst, MA 01003, USA}
\affiliation{David A. Dunlap Department of Astronomy and Astrophysics, University of Toronto, 50 St. George Street, Toronto, Ontario, M5S 3H4, Canada}
\email{j.antwidanso@utoronto.ca} 

\author[orcid=0000-0001-7549-5560]{Samantha C Berek}
\affiliation{Department of Astronomy, University of Massachusetts Amherst, 710 North Pleasant Street, Amherst, MA 01003, USA}
\email{sberek@umass.edu} 

\author[orcid=0000-0001-9002-3502]{Danilo Marchesini}
\affiliation{Department of Physics and Astronomy, Tufts University, 574 Boston Avenue, Suite 304, Medford, MA 02155, USA}
\email{danilo.marchesini@tufts.edu} 

\author[orcid=0009-0000-5385-8674]{Maya Merchant}
\affiliation{Department of Physics and Astronomy, York University, 4700 Keele St. Toronto, Ontario, M3J 1P3, Canada}
\email{merchm@yorku.ca}  

\author[orcid=0009-0009-2307-2350]{Katherine Myers}
\affiliation{Department of Physics and Astronomy, York University, 4700 Keele St. Toronto, Ontario, M3J 1P3, Canada}
\email{kjmyers@yorku.ca}

\begin{abstract}

JWST NIRCam spatially resolved photometry enables studies of galaxies' stellar populations and stellar masses on sub-kpc scales. We use JWST (CANUCS, Technicolor, JUMPS)  and HST data, covering wide- and medium-band UV-IR filters to study the stellar mass bias between spatially resolved and integrated photometry of $\sim$3000 galaxies at 1 $<z_{\mathrm{phot}}<$ 9. The 2D parameter maps reveal spatial variations in stellar mass and populations, showing that integrated measurements systematically underestimate stellar masses. We find that the stellar mass bias remains small for galaxies with specific star formation rate $\mathrm{sSFR} < 10^{\mathrm{-9}}$ yr$^{\mathrm{-1}}$, but increases for galaxies with $\mathrm{sSFR} > 10^{\mathrm{-8.5}}$ yr$^{\mathrm{-1}}$. For these high-sSFR galaxies, the median offset reaches $\Delta \log_{\mathrm{10}} (M_{\ast})$ $\sim$ -0.36 dex with the largest bias occurring in low-mass ($M_{\ast} \leq 10^{\mathrm{ 9.5}} M_{\odot}$) galaxies. These results are consistent with the outshining of older populations by recent star formation. High-sSFR galaxies also show spatial variations in their mass-to-light ratios ($M_{\ast}/L_{V}$): the stellar mass bias decreases with $\sigma[\log_{\mathrm{10}}(M_{\ast}/L_{V})]$ for galaxies with $10^{\mathrm{-8}} \leq \mathrm{sSFR}[\mathrm{yr}^{\mathrm{-1}}] < 10^{\mathrm{-9}}$, with the strongest trend for $10^{\mathrm{-7}} \leq \mathrm{sSFR}[\mathrm{yr}^{\mathrm{-1}}] < 10^{\mathrm{-8}}$, quantified by the Spearman correlation coefficient $\rho_{\rm S}=-0.37$. While medium bands are important for constraining stellar masses from integrated photometry, spatially resolved photometry reduces the impact of their absence, with dispersion offsets of $\Delta \log_{\mathrm{10}} (M_{\ast})$ = 0.33 dex for integrated photometry but 0.15 dex for resolved photometry when only wide-bands are included.

\end{abstract}



\section{Introduction}\label{sec: introduction}


A key aspect of understanding galaxy evolution is obtaining accurate measurements of fundamental galaxy properties, such as stellar mass, star formation, dust, and metallicity. Stellar mass, in particular, provides insight into the distribution and diversity of stellar populations within galaxies. Over time, a galaxy's stellar mass evolves through ongoing star formation, mergers, and episodic starbursts. Due to the processes that drive galaxy evolution, galaxies are often highly non-uniform, containing spatially distinct regions of young, bright stars alongside older, fainter populations. This complexity can cause integrated photometric measurements - all the light of a galaxy within a single aperture - to misrepresent the true stellar mass of a galaxy because they combine regions with different star formation histories and mass-to-light ratios. Spatially resolved multi-band photometry links global galaxy properties to the spatial distribution of stellar populations within galaxies. \cite{1999MNRAS.303..641A} was one of the first to look at spatially resolved multi-band photometry to investigate variations in stellar populations of galaxies at 0.4 $<$ $z$ $<$ 1, demonstrating the potential of internal color distributions to constrain relative ages and star formation histories (SFHs).

These internal variations are particularly important because integrated measurements can be affected by the outshining effect, in which recently formed, luminous O- and B-type stars dominate a galaxy’s integrated light, masking the contribution from older, less luminous stellar populations with higher mass-to-light ratios ($M_{\ast}/L_{V}$). This effect has been recognized for decades in integrated photometry (\citealp{1998AJ....115.1329S}; \citealp{2001ApJ...559..620P};  \citealp{2008MNRAS.386..715T}; \citealp{2010MNRAS.407..830M}; \citealp{2012MNRAS.422.3285P}; \citealp{2015MNRAS.452..235S}). Using pixel-based HST studies, \cite{2018MNRAS.476.1532S}, showed that galaxies affected by outshining exhibit significant spatial variation in their stellar populations, causing integrated SED fitting to underestimate stellar masses, particularly in high specific star formation rate ($\mathrm{sSFR} > 10^{\mathrm{-9}}$ yr$^{\mathrm{-1}}$) systems. Extending this analysis to galaxies up to $z \sim$ 2.5, they demonstrated that integrated SED fitting systematically underestimates stellar mass due to outshining. These findings highlighted the limitations of integrated photometric measurements and established the need for spatially resolved techniques capable of recovering mass hidden in less luminous stellar populations. Additional evidence for internal stellar population variations comes from spectroscopic analyses, such as \cite{2009ApJS..185..253G}, which revealed significant inhomogeneity in galaxy structures even when using spectroscopy. 
Resolved stellar mass maps have proven crucial for understanding how internal galaxy structure affects mass assembly 
(\citeauthor{2022ApJ...933...30T} \citeyear{2022ApJ...933...30T}; \citeyear{2024ApJ...964..177T}). 

On the contrary, \cite{2012ApJ...753..114W} did not find a significant  bias in the stellar mass between spatially integrated and resolved photometry for galaxies at 0.5 $<$ $z$ $<$ 2.5 using HST observations. \cite{2018MNRAS.476.1532S} discusses that the absence of an observed outshining bias in \cite{2012ApJ...753..114W} may be related to methodological choices, particularly the use of Voronoi binning and minimum-$\chi^2$ mass estimates, which reduced their sensitivity to spatial variations in stellar populations. Notably, when \cite{2012ApJ...753..114W} examined unbinned pixels, they actually found resolved masses higher by about 0.2 dex, but explained this by low signal-to-noise (S/N) pixels producing unreliable $M_{\ast}/L_{V}$ estimates. This highlights the challenge of maintaining sufficient spatial resolution in high-redshift HST observations, where Voronoi binning is often required to achieve adequate S/N. Maintaining high-S/N measurements while preserving spatial resolution therefore requires telescopes with substantially improved angular resolution.

This motivates the use of higher-resolution NIRCam observations from the James Webb Space Telescope (JWST) to extend resolved analyses to higher redshifts, where galaxies typically exhibit elevated sSFRs and are therefore prone to the outshining effect. JWST has opened new opportunities to study these effects at earlier cosmic times.
The extraordinary sensitivity and angular resolution of NIRCam (\citeauthor{2005SPIE.5904....1R}  \citeyear{2005SPIE.5904....1R}, \citeyear{2023PASP..135b8001R}) now allow spatially resolved analyses in the rest-frame optical for galaxies out to $z$ $\lesssim$ 6. 

The first demonstration of outshining with JWST was presented by \cite{2023ApJ...948..126G}, finding significant stellar mass underestimation of up to 1 dex in five lensed galaxies at 5 $<$ $z$ $<$ 9 in the SMACS0723 ERO field. These discrepancies were driven by compact regions of recent star formation that outshone older stellar populations in the outer parts of each galaxy. This picture is supported by \cite{2024ApJ...961L..21B}, who resolved multiple stellar components in the strongly lensed galaxy MACS1149-JD1 at $z$ = 9.1 and found that the bulk of the stellar mass resides in an older extended component, while ongoing star formation is concentrated in compact star-forming clumps. Similar behavior has been observed in other highly magnified systems, including a strongly lensed galaxy at $z$ = 6.072
\citep{2024A&A...686A..63G}  and the highly magnified "Cosmic Grapes" at $z$ = 6 \citep{2025NatAs...9.1553F}. At even earlier cosmic times, \cite{2025ApJ...995L..74B} resolved compact star-forming clumps on scales of $\leq$ 10 pc in the strongly lensed galaxy BulletArc-z11 at $z$ = 11.1, demonstrating that JWST combined with strong lensing can probe the internal structure of galaxies well into the epoch of reionization.

However, outshining is not universal. In contrast, more massive or red galaxies at $z$ $>$ 3 show minimal outshining, consistent with smoother, non-bursty star-formation histories (\citeauthor{2023ApJ...946L..16P} \citeyear{2023ApJ...946L..16P}; \citeauthor{2025MNRAS.539.2685L} \citeyear{2025MNRAS.539.2685L}). JWST has further revealed the complexity of internal stellar populations. \cite{2025arXiv250925363S} used NIRCam imaging to map the ages and masses of star-forming clumps at 0.5 $<$ $z$ $<$ 5, finding substantial kiloparsec variations in stellar population ages. 

Building on these early demonstrations, larger JWST surveys are extending resolved analyses beyond individual lensed systems. The CAnadian NIRISS Unbiased Cluster Survey (CANUCS; \citealp{2022PASP..134b5002W}) provides deep NIRISS and NIRCam observations of five strongly lensed galaxy cluster fields, enabling spatially resolved studies of high-redshift galaxies across a wider population. Beyond improved spatial resolution, JWST’s expanded wavelength coverage, including medium-band imaging, provides improved sampling of galaxy SEDs and enables more accurate characterization of stellar populations. Recent work highlights the importance of combining wide and medium bands for accurate photometric stellar mass estimates. \cite{2024ApJ...967L..17S} used JWST NIRCam observations from CANUCS and found that integrated stellar masses can differ systematically when medium bands are omitted, demonstrating that their inclusion yields more reliable stellar mass estimates. 

The choice of SFH parameterization and SED-fitting methodology also influences stellar mass recovery in spatially resolved analyses. \cite{2025MNRAS.542.2998H} used JWST/NIRCam imaging from the JADES Origins Field \citep{2026ApJS..283....6E} to investigate outshining in over 200 galaxies at z $\geq$ 4.5, finding that stellar masses can be systematically underestimated in low-mass galaxies with bursty SFHs. They further showed that more flexible SFH models, including double power-law and non-parametric continuity models, provide more consistent stellar mass estimates, although uncertainties remain due to SED-fitting degeneracies. In particular, when testing \texttt{Dense Basis} - a nonparametric SED fitting code - (\citealp{2017ApJ...838..127I}; \citealp{2019ApJ...879..116I}), they found improved agreement between integrated and resolved stellar masses compared to more restrictive SFH models, although low-mass galaxies still exhibited larger differences due to SFH priors rather than the data themselves. Complementing these observational studies, \cite{2024ApJ...961...73N} used cosmological simulations and mock SEDs to demonstrate that uncertainties in the stellar mass of high-redshift ($z >$ 7) galaxies can arise from the loss of early SFH information, as recent star formation outshines older stellar populations. In addition to SFH assumptions, spatial variations in dust attenuation can introduce further biases in integrated stellar mass estimates. \cite{2026arXiv260115963H} found using SED-fitting code \texttt{CIGALE} \citep{2019A&A...622A.103B} with parametric SFHs that resolved stellar masses are systematically higher than integrated estimates for galaxies at 3 $\leq$ $z$ $<$ 9, with the largest differences occurring in low-mass ($M_{\ast}$ $<$ 10$^8$ $M_{\odot}$) systems. They showed that these offsets are driven by the combined effects of outshining and spatially varying dust attenuation, which integrated SED fitting cannot fully capture.

Previous studies have provided important insights into the impact of outshining and the uncertainties in stellar mass estimates, but no study has combined non-parametric SFH modeling with spatially resolved SED fitting for a large galaxy sample with extensive medium-band coverage. Motivated by these findings, we apply Voronoi-binned spatially resolved SED-fitting, including both broadband and medium-bands, with \texttt{Dense Basis} - using a nonparametric SFH - to a sample of $\sim$3000 galaxies from the CANUCS / Technicolor (\citealp{2022PASP..134b5002W}; \citetalias{2026ApJS..282....3S} \citeyear{2026ApJS..282....3S}) and JWST Ultimate Medium-band Photometric Survey (JUMPS; Withers et al. in prep) deep field surveys.   

This work expands on previous work by measuring the difference between integrated and spatially resolved photometry to estimate the stellar mass for a large sample of galaxies at 1 $< z <$ 9 observed with JWST NIRCam. The paper is structured as follows. Section \ref{sec: data and sample selection} describes the data set; Section \ref{sec: method} the applied photometry, the \texttt{Voronoi} algorithm, and the SED fitting with the \texttt{Dense Basis}; Section \ref{sec: results} presents the results; Section \ref{sec: discussion} discusses the findings; and the summary with final conclusions is listed in Section \ref{sec: conclusion}. This work assumes a cosmology of $H_0$= 70 kms$^{-1}$Mpc$^{-1}$, $\Omega_m$ = 0.30, and $\Omega_{\Lambda}$ = 0.70. The magnitudes are in the AB system \citep{1983ApJ...266..713O}; flux $f_{\nu}$ is measured in nJy (10$^{-35}$ erg cm$^{-2}$ s$^{-1}$ Hz$^{-1}$) and AB$_{\nu}$ = -2.5 log$_{10}$(f$_{\nu}$/nJy) + 31.4. As well does this work uses the initial mass function (IMF) by \cite{2003PASP..115..763C}.

\section{Data \& Sample Selection}\label{sec: data and sample selection}
   \begin{table*}[hbt!]
        \centering
        \caption{List of all filters for each CANUCS lensing cluster (\textit{CLU}) and the parallel NIRCam Flanking (\textit{NCF}) field. JWST observations include NIRCam and NIRISS from the CANUCS, JUMPS, and Technicolor programs--the narrow bands are not included. HST data are drawn with instruments ACS/WFC, WFC3/UVIS, and WFC3/IR, spanning optical to near-infrared wavelengths. Detailed information on filter depths and coverage for each field can be found in (\citetalias{2026ApJS..282....3S} \citeyear{2026ApJS..282....3S}).}
    
        \footnotesize
        \setlength{\tabcolsep}{3pt}
        \renewcommand{\arraystretch}{0.9}
    
        \begin{tabular}{p{0.14\linewidth}p{0.2\linewidth}p{0.15\linewidth}p{0.14\linewidth}p{0.15\linewidth}p{0.14\linewidth}}
    
        \hline \hline
    
        \multirow{2}{*}{\centering \textbf{Field}} &
        \multicolumn{2}{c}{\textbf{JWST}} &
        \multicolumn{3}{c}{\textbf{HST}} \\
    
        & \textit{NIRCam}
        & \textit{NIRISS}
        & \textit{ACS/WFC}
        & \textit{WFC3/UVIS}
        & \textit{WFC3/IR} \\
    
        \hline \hline 
    
        \parbox[t]{\linewidth}{%
        \vspace{0.4\baselineskip}
        A370-CLU\\
        Filters: 26%
        }
        
        &
        \vspace{-1.5\baselineskip}
        \begin{tabular}[c]{@{}l@{}}
        F090W, F115W, F150W\\
        F200W, F277W, F356W\\
        F360M, F410M, F430M\\
        F444W, F460M, F480M
        \end{tabular}
        &
        \begin{tabular}[c]{@{}l@{}}
        F090WN, F115WN\\
        F150WN, F200WN
        \end{tabular}
        &
        \begin{tabular}[c]{@{}l@{}}
        F435W, F475W\\
        F606W, F625W\\
        F814W
        \end{tabular}
        &
        —
        &
        \begin{tabular}[c]{@{}l@{}}
        F105W, F110W\\
        F125W, F140W\\
        F160W
        \end{tabular}
        \\
    
        \hline
    
        \parbox[t]{\linewidth}{%
        \vspace{0.4\baselineskip}
        A370-NCF\\
        Filters: 27%
        }
        &
        \vspace{-1.5\baselineskip}
        \begin{tabular}[c]{@{}l@{}}
        F070W, F090W, F115W\\
        F140M, F150W, F162M\\
        F182M, F200W, F210M\\
        F250M, F277W, F300M\\
        F335M, F356W, F360M\\
        F410M, F430M, F444W\\
        F460M, F480M
        \end{tabular}
        &
        —
        &
        \begin{tabular}[c]{@{}l@{}}
        F435W, F475W\\
        F606W, F625W\\
        F814W
        \end{tabular}
        &
        —
        &
        \begin{tabular}[c]{@{}l@{}}
        F105W, F110W\\
        F125W, F140W\\
        F160W
        \end{tabular}
        \\
    
        \hline
    
        \begin{tabular}[c]{@{}l@{}}
        MACS0416-CLU\\
        Filters: 24
        \end{tabular}
        &
        \vspace{-1.5\baselineskip}
        \begin{tabular}[c]{@{}l@{}}
        F090W, F115W, F150W\\
        F200W, F277W, F356W\\
        F360M, F410M, F430M\\
        F444W, F460M, F480M
        \end{tabular}
        &
        \begin{tabular}[c]{@{}l@{}}
        F090WN, F115WN\\
        F150WN, F200WN
        \end{tabular}
        &
        \begin{tabular}[c]{@{}l@{}}
        F435W, F475W\\
        F606W, F625W\\
        F814W
        \end{tabular}
        &
        —
        &
        \begin{tabular}[c]{@{}l@{}}
        F105W, F110W\\
        F125W, F140W\\
        F160W
        \end{tabular}
        \\
    
        \hline

        \parbox[t]{\linewidth}{%
        \vspace{0.4\baselineskip}
        MACS0416-NCF\\
        Filters: 27%
        }
        &
        \vspace{-1.5\baselineskip}
        \begin{tabular}[c]{@{}l@{}}
        F070W, F090W, F115W\\
        F140M, F150W, F162M\\
        F182M, F200W, F210M\\
        F250M, F277W, F300M\\
        F335M, F356W, F360M\\
        F410M, F430M, F444W\\
        F460M, F480M
        \end{tabular}
        &
        —
        &
        \begin{tabular}[c]{@{}l@{}}
        F435W, F475W\\
        F606W, F625W\\
        F814W
        \end{tabular}
        &
        —
        &
        \begin{tabular}[c]{@{}l@{}}
        F105W, F110W\\
        F125W, F140W\\
        F160W
        \end{tabular}
        \\
    
        \hline
    
        \parbox[t]{\linewidth}{%
        \vspace{0.4\baselineskip}
        MACS0417-CLU\\
        Filters: 18%
        }
        &
        \vspace{-1.5\baselineskip}
        \begin{tabular}[c]{@{}l@{}}
        F090W, F115W, F150W\\
        F200W, F277W, F356W\\
        F410M, F444W
        \end{tabular}
        &
        \begin{tabular}[c]{@{}l@{}}
        F115WN, F150WN\\
        F200WN
        \end{tabular}
        &
        \begin{tabular}[c]{@{}l@{}}
        F435W, F475W\\
        F606W, F625W\\
        F814W
        \end{tabular}
        &
        —
        &
        \begin{tabular}[c]{@{}l@{}}
        F105W, F110W\\
        F125W, F140W\\
        F160W
        \end{tabular}
        \\
    
        \hline
    
        \parbox[t]{\linewidth}{%
        \vspace{0.4\baselineskip}
        MACS0417-NCF\\
        Filters: 16%
        }
        &
        \vspace{-1.5\baselineskip}
        \begin{tabular}[c]{@{}l@{}}
        F090W, F115W, F140M\\
        F150W, F162M, F182M\\
        F210M, F250M, F277W\\
        F300M, F335M, F360M\\
        F410M, F444M
        \end{tabular}
        &
        —
        &
        —
        &
        \begin{tabular}[c]{@{}l@{}}
        F438WU, F606WU
        \end{tabular}
        &
        —
        \\
    
        \hline
    
        \parbox[t]{\linewidth}{%
        \vspace{0.4\baselineskip}
        MACS1149-CLU\\
        Filters: 24%
        }
        &
        \vspace{-1.5\baselineskip}
        \begin{tabular}[c]{@{}l@{}}
        F090W, F115W, F150W\\
        F200W, F277W, F356W\\
        F360M, F410M, F430M\\
        F444W, F460M, F480M
        \end{tabular}
        &
        \begin{tabular}[c]{@{}l@{}}
        F090WN, F115WN\\
        F150WN, F200WN
        \end{tabular}
        &
        \begin{tabular}[c]{@{}l@{}}
        F435W, F475W\\
        F606W, F625W\\
        F814W
        \end{tabular}
        &
        —
        &
        \begin{tabular}[c]{@{}l@{}}
        F105W, F110W\\
        F125W, F140W\\
        F160W
        \end{tabular}
        \\
    
        \hline
    
        \parbox[t]{\linewidth}{%
        \vspace{0.4\baselineskip}
        MACS1149-NCF\\
        Filters: 25%
        }
        &
        \vspace{-1.5\baselineskip}
        \begin{tabular}[c]{@{}l@{}}
        F070W, F090W, F115W\\
        F140M, F150W, F182M\\
        F200W, F210M, F277W\\
        F300M, F335M, F356W\\
        F360M, F410M, F430M\\
        F444W, F460M, F480M
        \end{tabular}
        &
        —
        &
        \begin{tabular}[c]{@{}l@{}}
        F435W, F475W\\
        F606W, F625W\\
        F814W
        \end{tabular}
        &
        —
        &
        \begin{tabular}[c]{@{}l@{}}
        F105W, F110W\\
        F125W, F140W\\
        F160W
        \end{tabular}
        \\
    
        \hline
    
        \parbox[t]{\linewidth}{%
        \vspace{0.4\baselineskip}
        MACS1423-CLU\\
        Filters: 19%
        }
        &
        \vspace{-1.5\baselineskip}
        \begin{tabular}[c]{@{}l@{}}
        F090W, F115W, F150W\\
        F200W, F277W, F356W\\
        F410M, F444W
        \end{tabular}
        &
        \begin{tabular}[c]{@{}l@{}}
        F115WN, F150WN\\
        F200WN
        \end{tabular}
        &
        \begin{tabular}[c]{@{}l@{}}
        F435W, F475W\\
        F606W, F625W\\
        F814W
        \end{tabular}
        &
        —
        &
        \begin{tabular}[c]{@{}l@{}}
        F105W, F110W\\
        F125W, F140W\\
        F160W
        \end{tabular}
        \\
    
        \hline
    
        \parbox[t]{\linewidth}{%
        \vspace{0.4\baselineskip}
        MACS1423-NCF\\
        Filters: 18%
        }
        &
        \vspace{-1.5\baselineskip}
        \begin{tabular}[c]{@{}l@{}}
        F090W, F115W, F140M\\
        F150W, F162M, F182M\\
        F210M, F250M, F277W\\
        F300M, F335M, F360M\\
        F410M, F444W
        \end{tabular}
        &
        —
        &
        —
        &
        \begin{tabular}[c]{@{}l@{}}
        F438WU, F606WU
        \end{tabular}
        &
        \begin{tabular}[c]{@{}l@{}}
        F105W, F110W\\
        F125W, F140W\\
        F160W
        \end{tabular}
        \\
    
        \hline
    
        \end{tabular} \label{tab:clusters_and_filters}
    \end{table*}

    \subsection{\textit{Imaging: HST \& JWST}}\label{subsec: jwst hst}

    This work uses galaxies in the background of the five CANUCS lensing clusters \citep{2022PASP..134b5002W}: Abell 370 at $z$ = 0.375 (A370, \citealp{2017ApJ...837...97L}), MACS J0416.1-2403 at $z$ = 0.395 (MACS0416, \citealp{2017ApJ...837...97L}), MACS J0417.5-1154 at $z$ = 0.443 (MACS0417, \citealp{2012ApJS..199...25P}), MACS J1149.5+2223 at $z$ = 0.543 (MACS1149, \citealp{2017ApJ...837...97L}), and MACS J1423.8+2404 at $z$ = 0.545 (MACS1423, \citealp{2020ApJ...889..189S}). These clusters and their flanking fields were observed with JWST/NIRCam as part of the NIRISS GTO Program PID 1208, CANUCS and Cycle 2 GO program PID 3362 Technicolor (\citeauthor{2022PASP..134b5002W} \citeyear{2022PASP..134b5002W}; \citetalias{2026ApJS..282....3S} \citeyear{2026ApJS..282....3S}), and the Cycle 3 GO program PID 5890 JUMPS (Withers et al. in prep). Photometry for all cluster and background sources of CANUCS and Technicolor is provided in the CANUCS Data Release 1 (DR1) catalog (\citetalias{2026ApJS..282....3S} \citeyear{2026ApJS..282....3S}). Although not included in DR1, JUMPS was processed using an identical imaging and photometry pipeline.
    \begin{figure*}
        \centering
        \includegraphics[width=\textwidth]{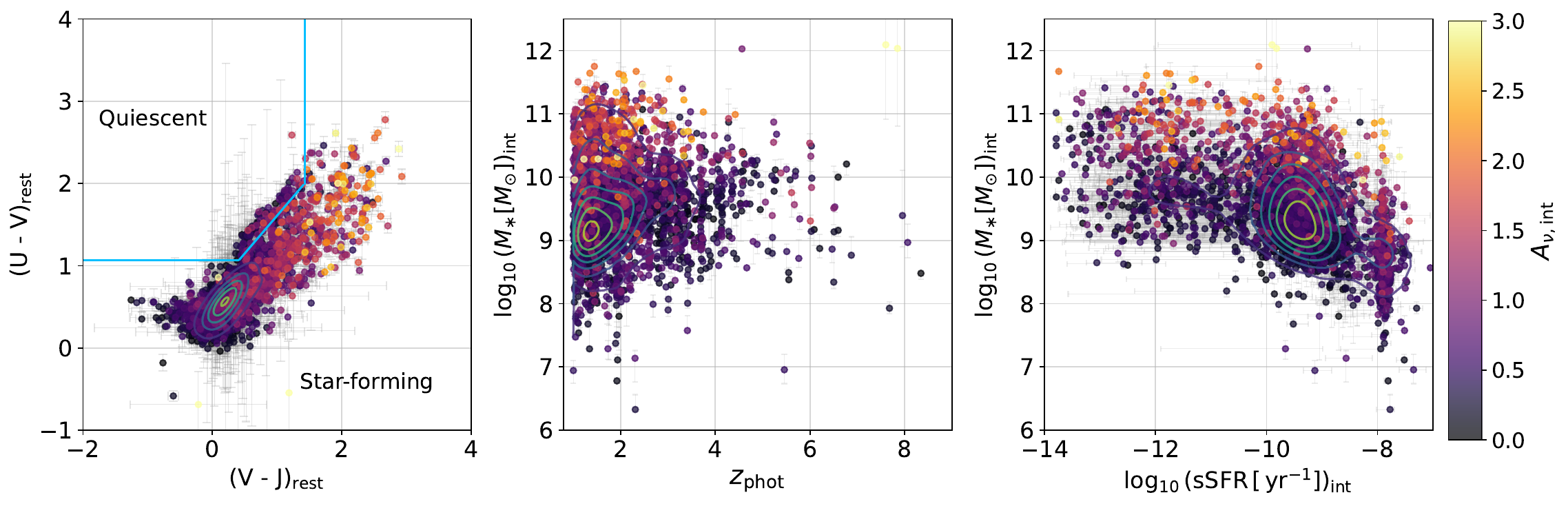}
        \hfill
        \caption{The total sample consists of 2980 galaxies selected from the clusters A370, MACS0416, MACS0417, MACS1149, and MACS1423 along with their flanking fields, as part of the CANUCS, Technicolor, and JUMPS programs. Galaxies were selected based on an integrated SNR$_{\mathrm{F444W}}$ $>$ 50 and a photometric redshift in the range 1 $<$ $z_{\mathrm{phot}}$ $<$ 9. UVJ colors and $z_{\mathrm{phot}}$ are calculated from integrated photometry using \texttt{EAzY}, while the stellar mass $\log_{10}(M_{\ast}[M_{\odot}])$, specific star formation rate $\log_{10}(\mathrm{sSFR}[yr^{-1}])$, and dust attenuation $A_V$ were estimated using \texttt{Dense Basis}. The stellar masses from the \textit{CLU} fields are corrected for lensing magnification using the factors provided by the DR1 catalog. The left panel displays the position of galaxies in the UVJ color-color diagram, where the UVJ boundaries were determined by \cite{2023ApJ...943..166A}. The middle panel shows the integrated $\log_{10}(M_{\ast})$ distribution across $z_{\mathrm{phot}}$, and the right panel shows the $\log_{10}(M_{\ast})$-$\log_{10}$(sSFR) relation. The color bar, applied in all panels, represents $A_V$. From the total sample, 116 are UVJ-selected quiescent galaxies and 153 galaxies are at $z_{\mathrm{phot}}$ $\geq$ 4.}
        \label{fig: samples 3 panel}
    \end{figure*}
 
    The CANUCS survey includes imaging from JWST NIRCam and NIRISS. NIRCam has two pointings, one on the cluster center \textit{CLU} and on the NIRCam flanking field \textit{NCF}. The \textit{CLU} fields are affected by gravitational lensing; the lensing models are discussed for A370 in \cite{2024ApJ...973...77G}, MACS0416 in \cite{2025A&A...696A..15R}, MACS0417 in (Desprez et al. in prep.), MACS1149 in (Rihtar\v{s}i\v{c} et al. in prep.), and MACS1423 in (Desprez et al. in prep.). Stellar parameters of galaxies in the \textit{CLU} field are corrected for lensing magnification. All fields also have complementary HST imaging from ACS/WFC, WFC3/UVIS, or WFC3/IR (\citealp{2012ApJS..199...25P}; \citealp{2017ApJ...837...97L}; \citealp{2020ApJ...889..189S}; \citealp{2020ApJS..247...64S}). The full set of available filters for each cluster and its flanking field is summarised in \Cref{tab:clusters_and_filters}. Two narrow-band filters from Technicolor, F164N and F187N, are excluded from our analysis because they primarily trace rare extreme emission line sources and do not provide useful additional constraints for SED fitting regarding the scope of this paper. The images used in this paper were all convolved to the F444W PSF. Also note that the photometric redshifts and lensing magnification factors from DR1 are calculated within an aperture of 0.3". Further details for CANUCS and Technicolor on the survey design, data reduction, source detection, filter coverage, and effective depths (3$\sigma\sim$ 29 mag for most filters) are provided in DR1.

    \begin{figure*}
        \centering
        \includegraphics[width=0.49\textwidth]{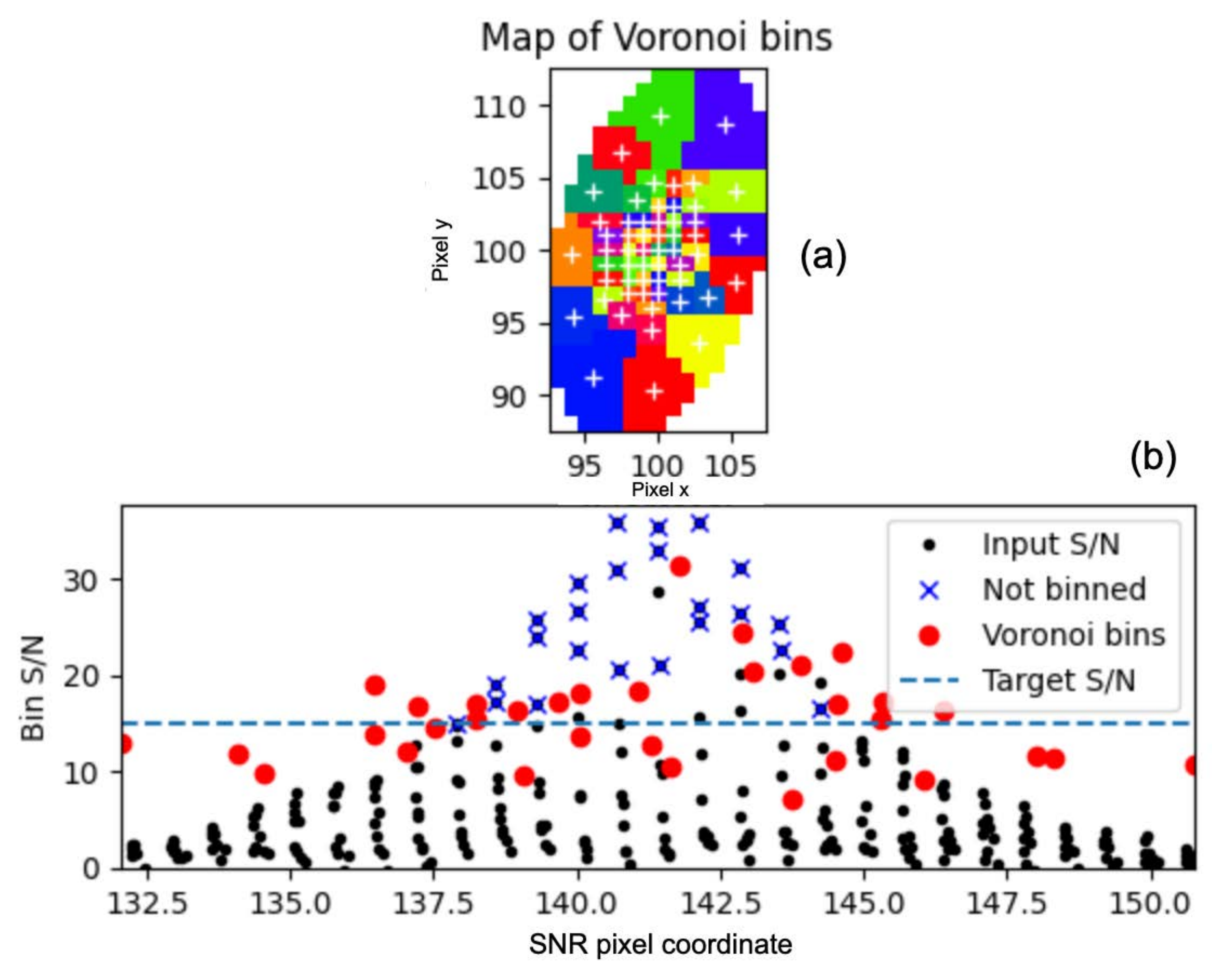}
        \hfill
        \includegraphics[width=0.49\textwidth]{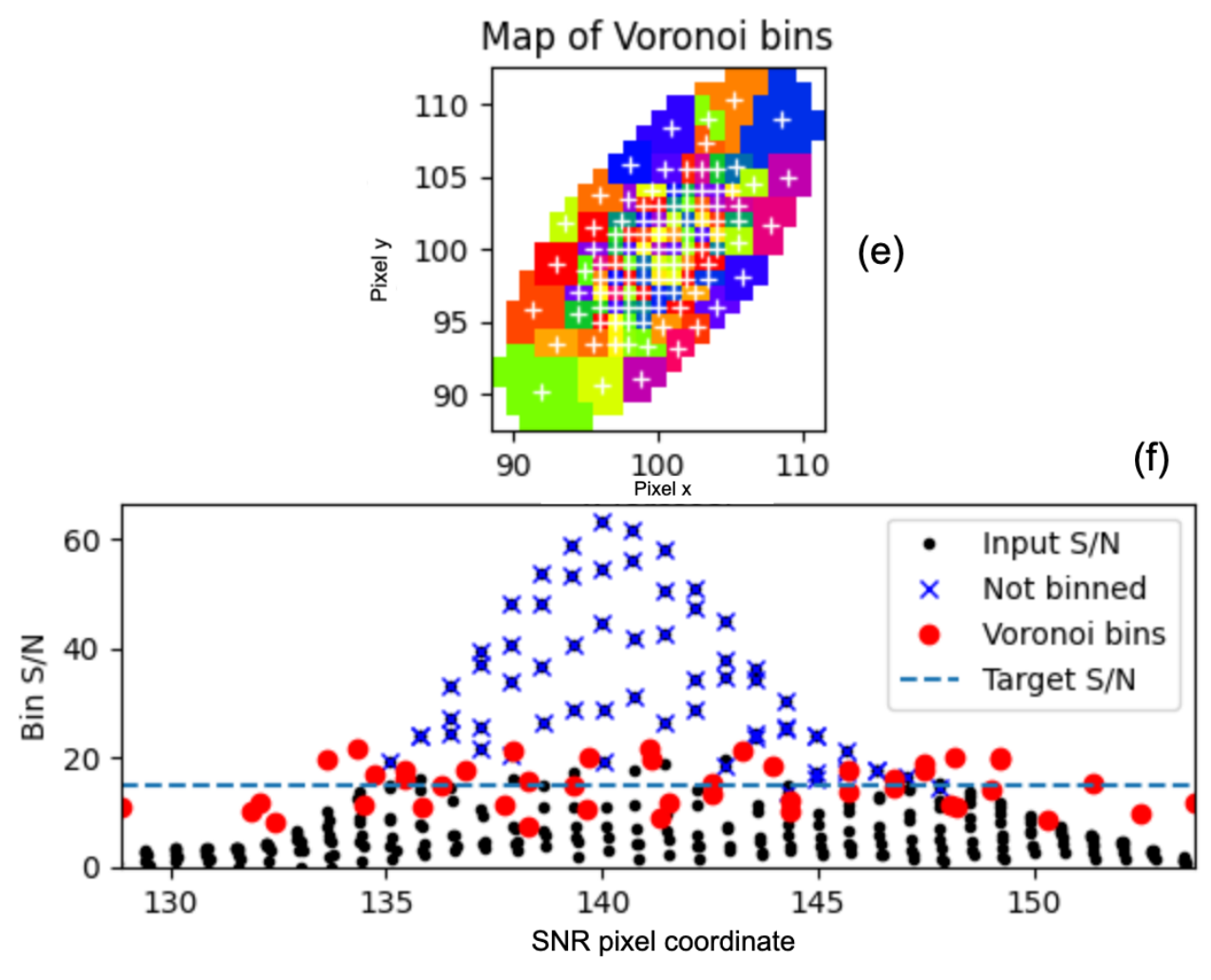}
        \\[0.5em]
        \makebox[\textwidth][c]{%
            \begin{minipage}{0.49\textwidth}
                \centering
                \includegraphics[width=\textwidth]{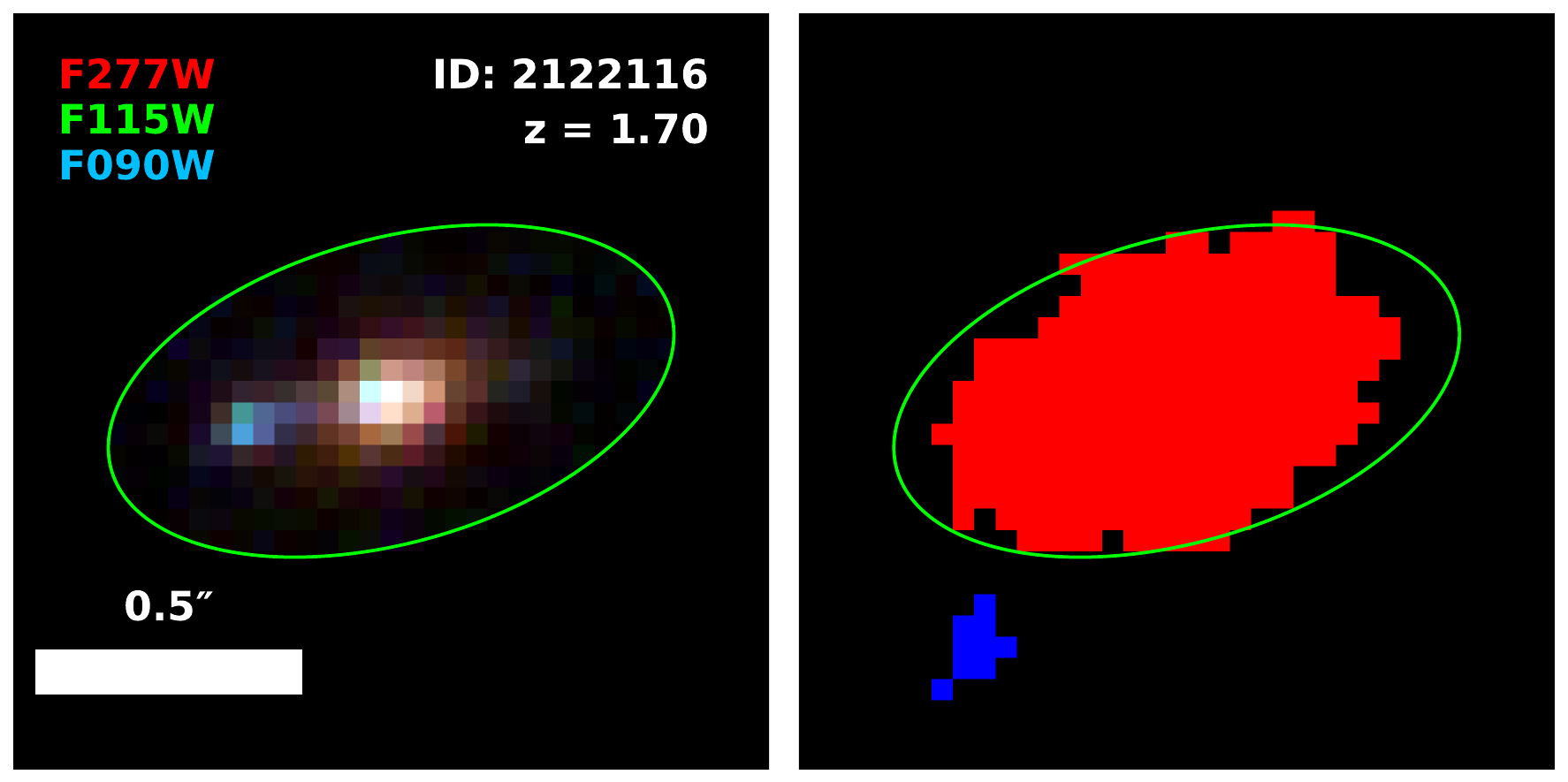}\\[-0.3em]
                \makebox[\textwidth]{%
                    \hfill (c) \hfill\hfill (d) \hfill
                }
            \end{minipage}
            \hfill
            \begin{minipage}{0.49\textwidth}
                \centering
                \includegraphics[width=\textwidth]{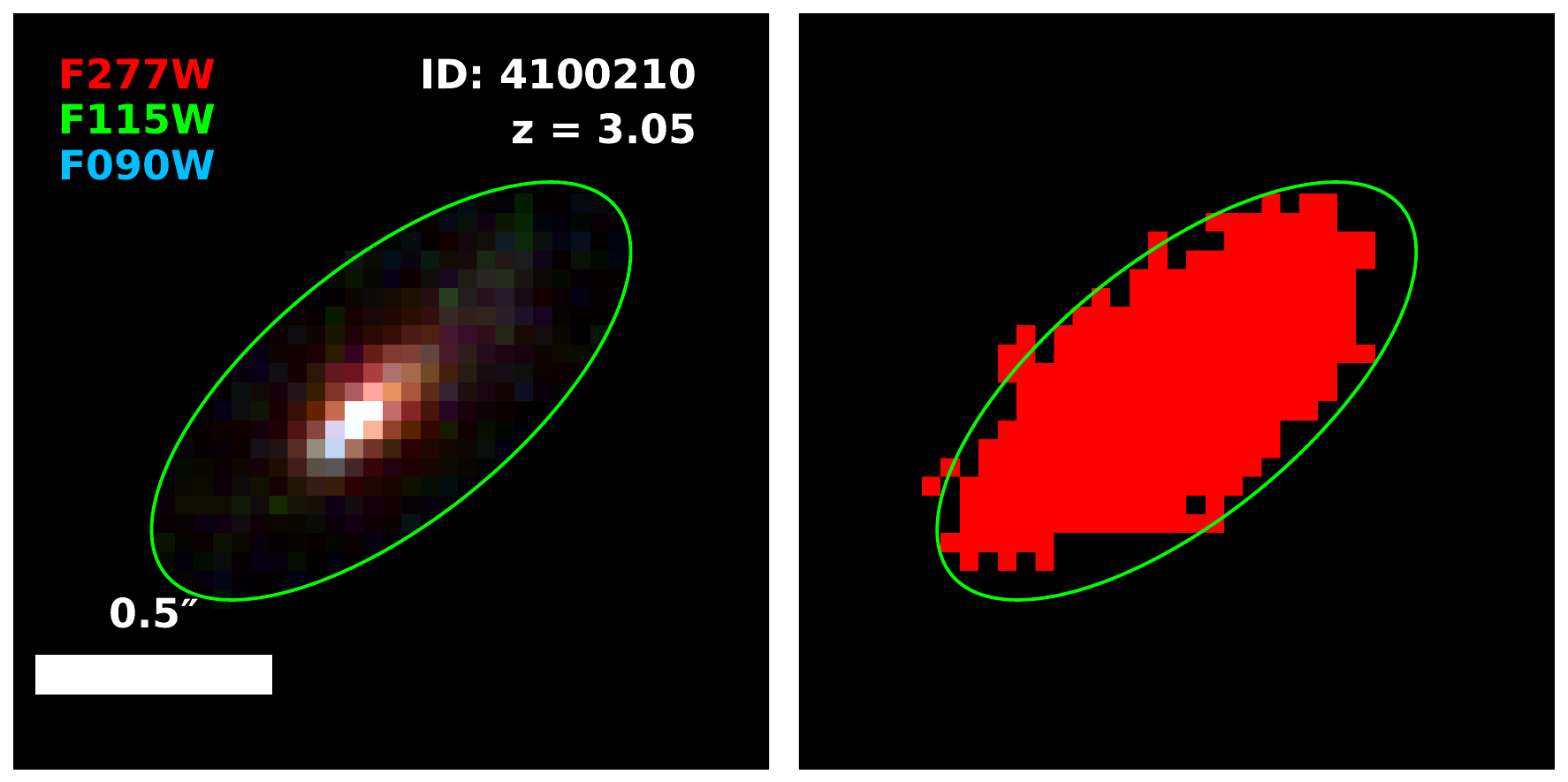}\\[-0.3em]
                \makebox[\textwidth]{%
                    \hfill (g) \hfill\hfill (h) \hfill
                }
            \end{minipage}
        }
    
        \caption{
        (a) Voronoi map of galaxy 2122116.
        (b) SNR of individual pixels (black dots) and Voronoi bins (red dots) in F444W as a function of the pixel coordinates used by the Voronoi algorithm. Unbinned pixels shown as blue crosses on top of black dots, for galaxy 2122116. The dashed blue line is the target SNR of 15. 
        (c) RGB image of galaxy 2122116.
        (d) Segmentation map of galaxy 2122116.
        (e) Voronoi map of galaxy 4100210.
        (f) SNR of pixels (black dots) and Voronoi bins (red dots) in F444W as a function of the pixel coordinates used by the Voronoi algorithm. Unbinned pixels shown as blue crosses on top of black dots, for galaxy 4100210. The dashed blue line is the target SNR of 15.
        (g) RGB image of galaxy 4100210.
        (h) Segmentation map of galaxy 4100210.
        Red indicates the segmentation of the target object and blue indicates neighbouring target IDs, which are masked if they fall within the Kron aperture and subsequently filled using mirrored pixels. Panels a, b, e, and f are generated by Voronoi.}
        \label{fig: voronoi_rgb_segmap_examples}
    \end{figure*}

    \subsection{Sample selection}\label{subsec: target selection}
    In this paper, we apply a set of selection criteria to all available targets in the CANUCS cluster and flanking fields. The sample is defined using DR1 catalogue flags together with two additional requirements tailored to this study. We start with removing point sources (\texttt{FLAG\_POINTSRC}, \texttt{FLAG\_GAIA}), bright cluster galaxies (\texttt{FLAG\_BCG}), objects with substantial contamination from cluster halos (\texttt{FLAG\_HIGHBCGCONTAMI}), and sources not detected in HST ACS (\texttt{FLAG\_ACS}) and JWST NIRCam (\texttt{FLAG\_NIRCAM}). A full description of these flags is provided in Table 7 of DR1.
    
    Galaxies are then selected based on their total signal-to-noise ratio (SNR) in the reddest wide-band filter, F444W, and on their photometric redshift $z_{\mathrm{phot}}$ derived using \texttt{EAzY} (\citealp{2008ApJ...686.1503B}; as provided in DR1). The adopted criteria are SNR$_{\mathrm{F444W}} > 50$ and $1 < z_{\mathrm{phot}} < 9$. Galaxies with SNR$_{\mathrm{F444W}} <$ 50 are excluded because Voronoi tessellation would produce too few spatial bins, which means there would be  no meaningful difference between integrated and resolved photometry. Objects at $z_{\mathrm{phot}} < 1$ are excluded because key SED features that constrain stellar mass estimates, such as Balmer/4000 $\AA$ break and the rest-frame optical continuum, would primarily be covered by HST bands rather than JWST/NIRCam, resulting in less uniform wavelength coverage compared with the higher-redshift sample. We also omit galaxies at $z_{\mathrm{phot}} > 9$ because key rest-frame optical features that constrain stellar population properties, including the Balmer/4000 $\AA$ break region and optical continuum, shift beyond NIRCam coverage into MIRI, which is not available for this data set. These criteria are designed to ensure reliable Voronoi tessellation and spatially resolved photometry, enabling a robust comparison between integrated and resolved stellar mass estimates. Including objects outside this redshift range would introduce redshift-dependent differences in the wavelength coverage used to constrain stellar populations, potentially leading to systematic differences in the inferred stellar masses. 
    
    \Cref{fig: samples 3 panel} shows the selected samples, along with their \texttt{EAzY}-derived rest-frame colors from DR1 and integrated photometric properties estimated from SED fitting with \texttt{Dense Basis} (see Section \ref{subsec: method dense basis} for priors and settings of \texttt{Dense Basis}). The CANUCS-calibrated boundaries between star-forming and quiescent galaxies in the UVJ diagram are defined by the following equations \citep{2023ApJ...943..166A}:
    \begin{equation}
        \begin{aligned}
             1 < z_{\mathrm{phot}} < 9: & \hspace*{0.5cm}  (U - V) > 1.07, \hspace*{0.2cm} (V - J) < 1.6 \\
                                & \hspace*{0.5cm}  (U-V) > (V-J) \times 0.92 + 0.69.
        \end{aligned}
    \end{equation}

    The left panel of \Cref{fig: samples 3 panel} shows the galaxy colors, with most galaxies classified as blue star-forming. The middle panel shows the integrated stellar mass as a function of photometric redshift $z_{\mathrm{phot}}$, and the right panel presents the stellar mass versus sSFR. The stellar masses from the \textit{CLU} fields are corrected for lensing magnification using the factors provided by the DR1 catalog. Approximately half of the sample has $\log_{10}(M_\ast/M_\odot) < 9.5$, which will be used later to split the sample, and 116 are classified as quiescent galaxies. A subset of low-mass galaxies with very low-sSFR also exhibits minimal dust attenuation. Among the sample of 2980, a total of 153 galaxies are at $z_{\mathrm{phot}} > 4.5$ and 116 are UVJ-selected quiescent galaxies.

\begin{table}\centering
\caption[]{Summary of the number of galaxies selected at each stage of the selection procedure for the \textit{CLU} and \textit{NCF} samples in each cluster field. The columns show the number of objects meeting the initial criteria ($N_{\mathrm{criteria}}$), those selected through the Voronoi method ($N_{\mathrm{Voronoi}}$), and the number of objects with valid integrated and resolved SED-fitting outputs ($N_{\mathrm{DB}}$), and the final sample size ($N_{\mathrm{final}}$).}
\label{tab: nr of galaxies}

\begin{tabular}{lp{.5cm}|p{.1cm}p{.1cm}p{.1cm}p{.1cm}}
\hline
\textbf{Field} & &
\multicolumn{1}{l}{$N_{\mathrm{criteria}}$} &
\multicolumn{1}{l}{$N_{\mathrm{Voronoi}}$} &
\multicolumn{1}{l}{$N_{\mathrm{DB}}$} &
\multicolumn{1}{l}{$N_{\mathrm{final}}$} \\
\hline \hline

A370 &
{CLU \newline NCF} &
{150 \newline 324} &
{149 \newline 324} &
{147 \newline 324} &
{145 \newline 324} \\
\hline

MACS0416 &
{CLU \newline NCF} &
{266 \newline 268} &
{266 \newline 268} &
{266 \newline 268} &
{265 \newline 268} \\
\hline

MACS0417 &
{CLU \newline NCF} &
{234 \newline 503} &
{234 \newline 503} &
{234 \newline 503} &
{231 \newline 497} \\
\hline

MACS1149 &
{CLU \newline NCF} &
{304 \newline 379} &
{304 \newline 379} &
{304 \newline 379} &
{301 \newline 379} \\
\hline

MACS1423 &
{CLU \newline NCF} &
{278 \newline 353} &
{278 \newline 353} &
{278 \newline 353} &
{278 \newline 352} \\
\hline

\end{tabular}
\end{table}
    
\section{Spatially Resolved Photometry \& SED-fitting}\label{sec: method}

\begin{figure*}
    \centering

    \includegraphics[width=\textwidth]{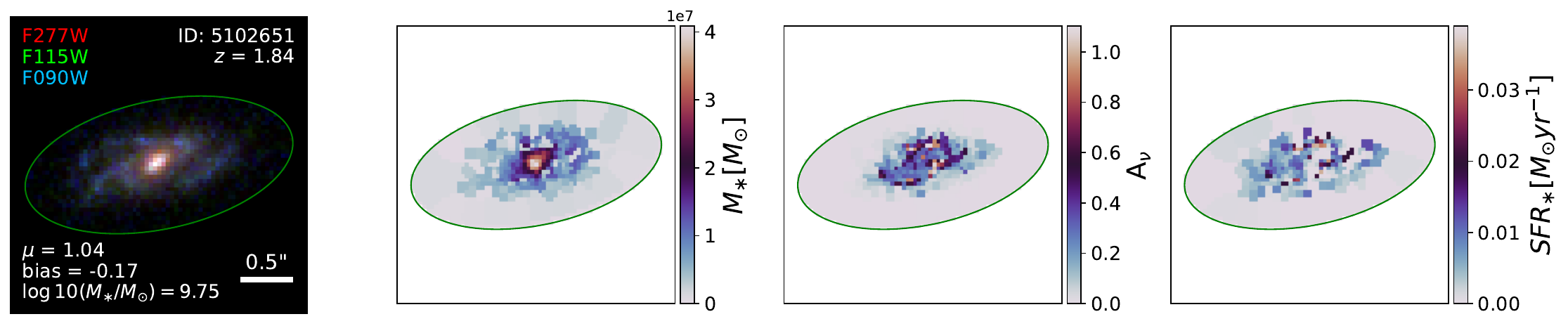}

    \includegraphics[width=\textwidth]{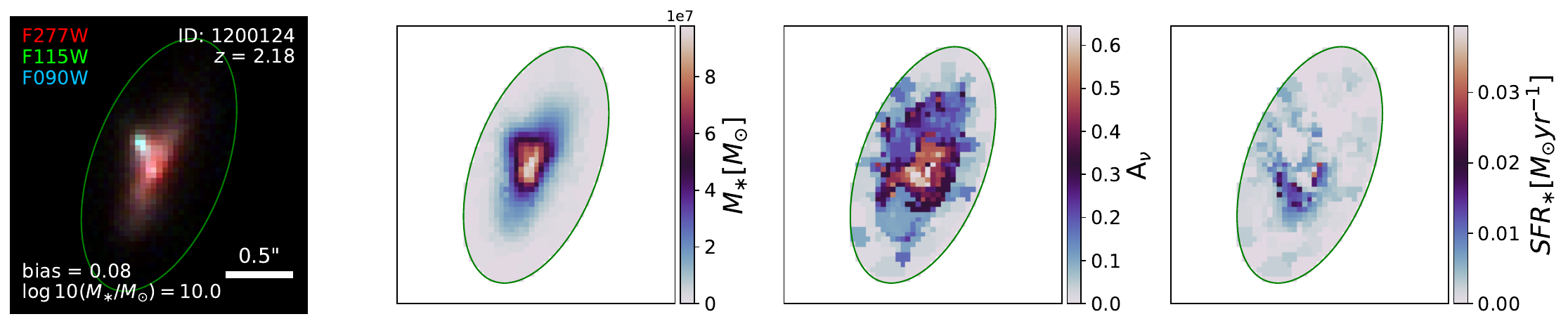}

    \includegraphics[width=\textwidth]{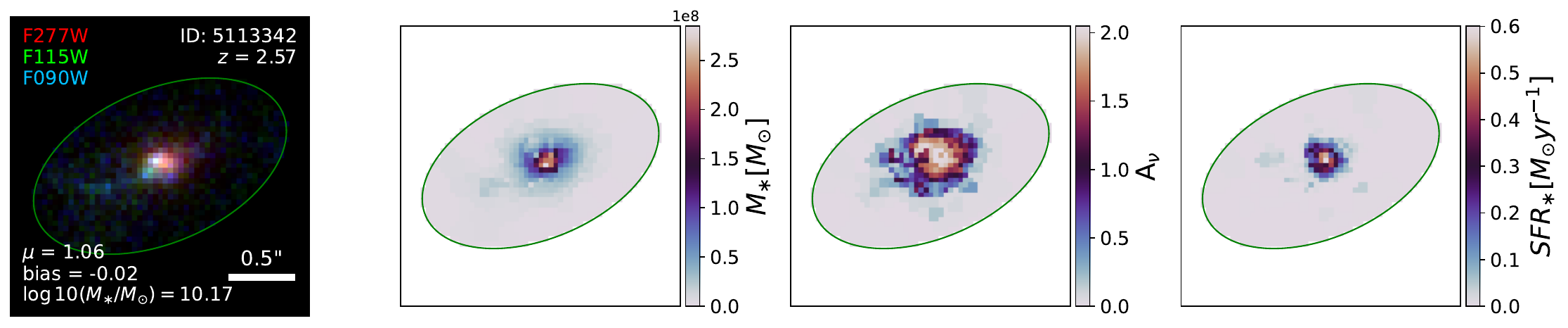}

    \includegraphics[width=\textwidth]{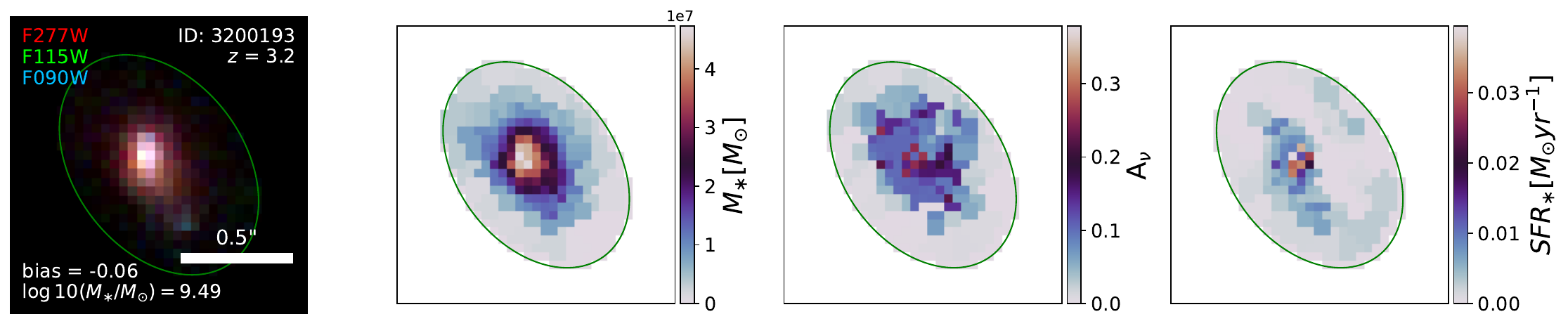}

    \includegraphics[width=\textwidth]{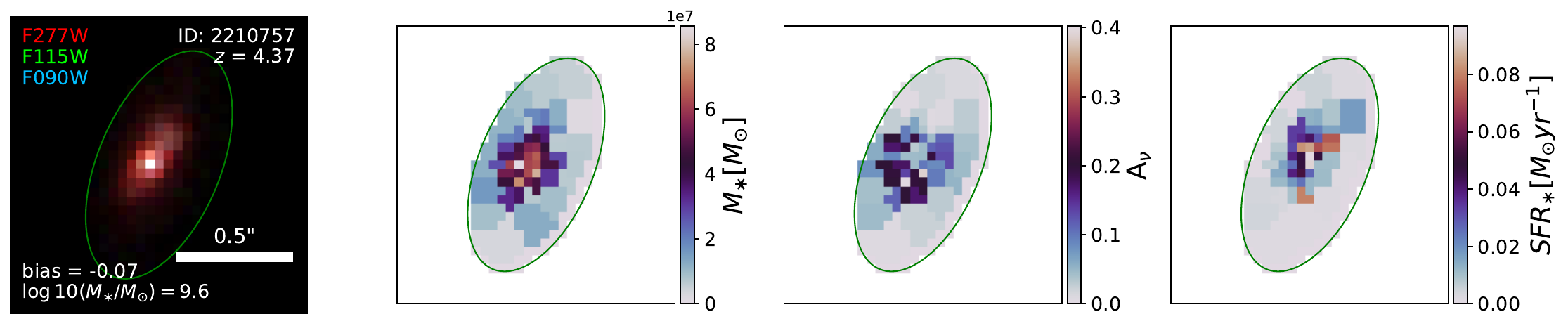}

    \includegraphics[width=\textwidth]{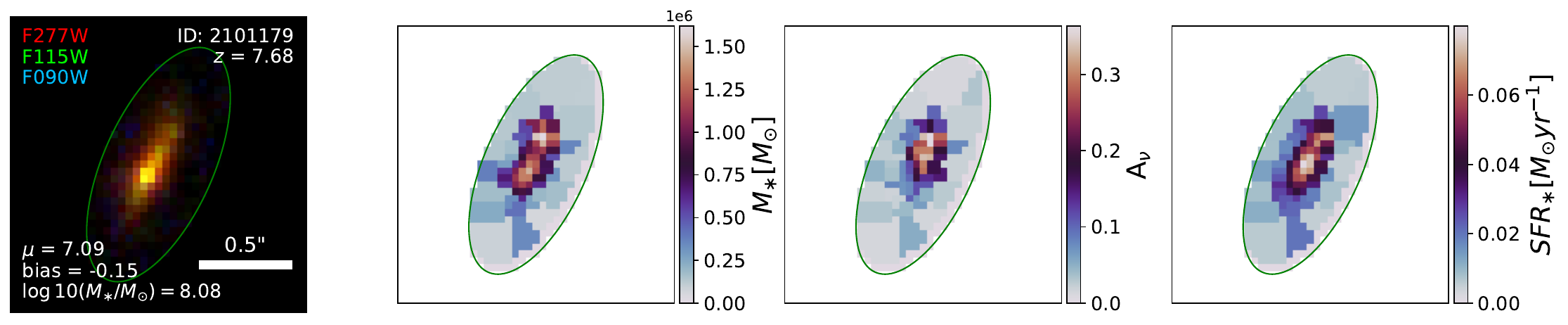}

    \caption{
        Galaxies with a small stellar mass bias. RGB and spatially resolved maps: stellar mass ($M_{\ast}$), dust attenuation ($A_{\nu}$), and star formation rate (SFR$_{\ast}$) of each pixel.
    }
    \label{fig: rgb and 2d maps small bias}
\end{figure*}

\begin{figure*}
    \centering
    \includegraphics[width=\textwidth]{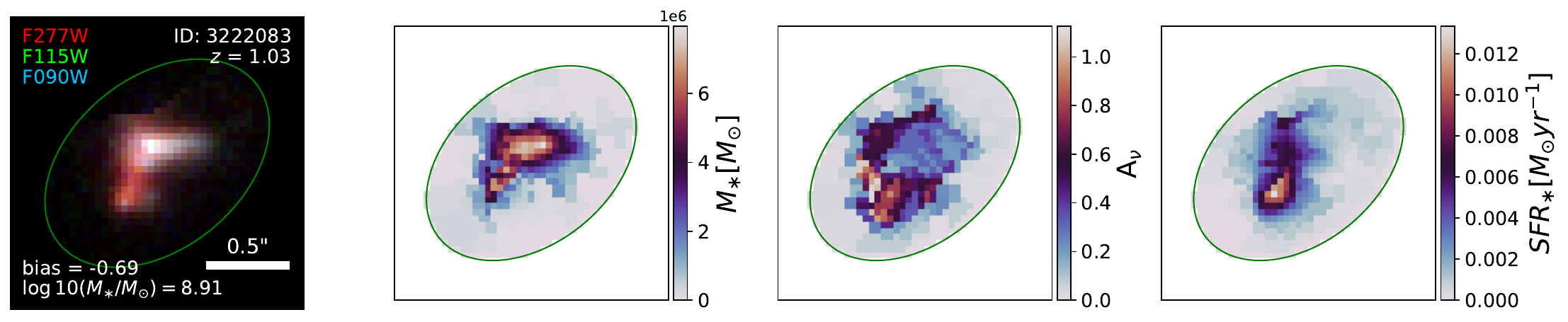}

    \includegraphics[width=\textwidth]{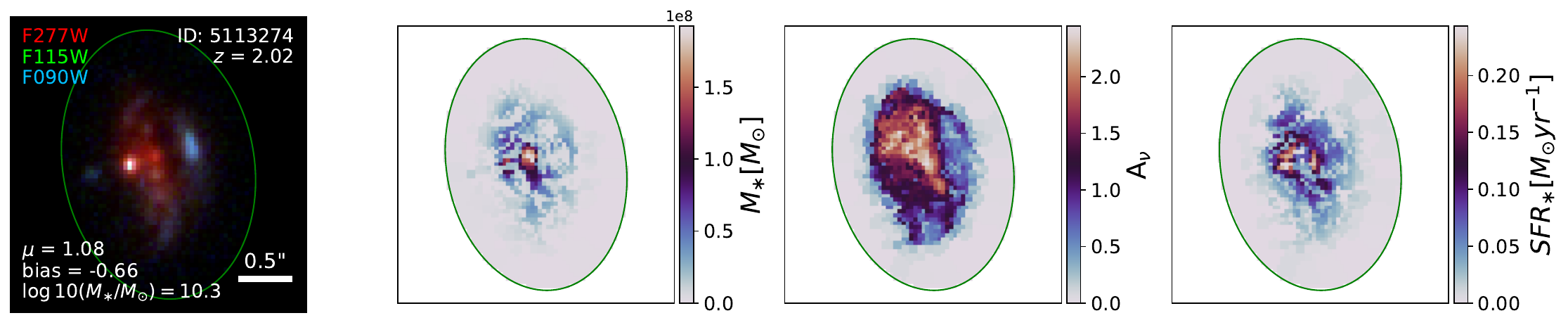}

    \includegraphics[width=\textwidth]{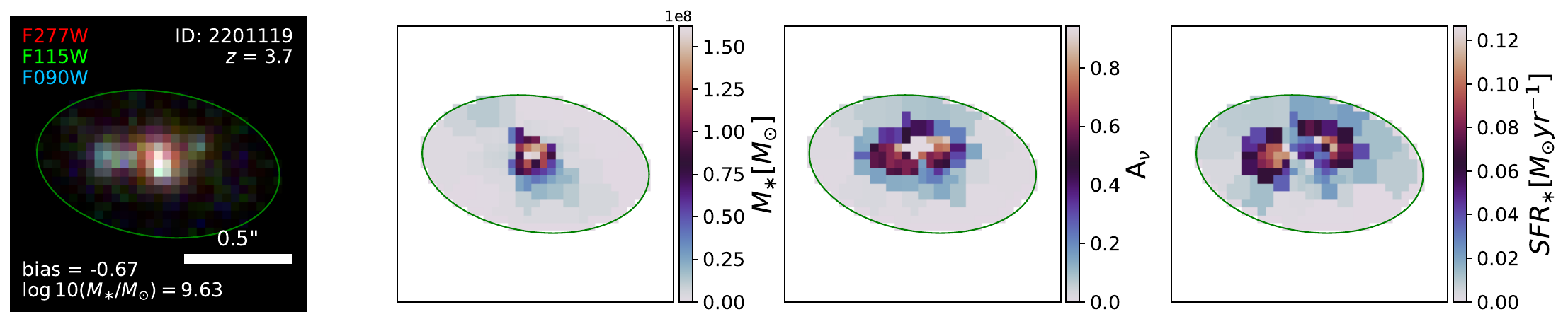}

    \includegraphics[width=\textwidth]{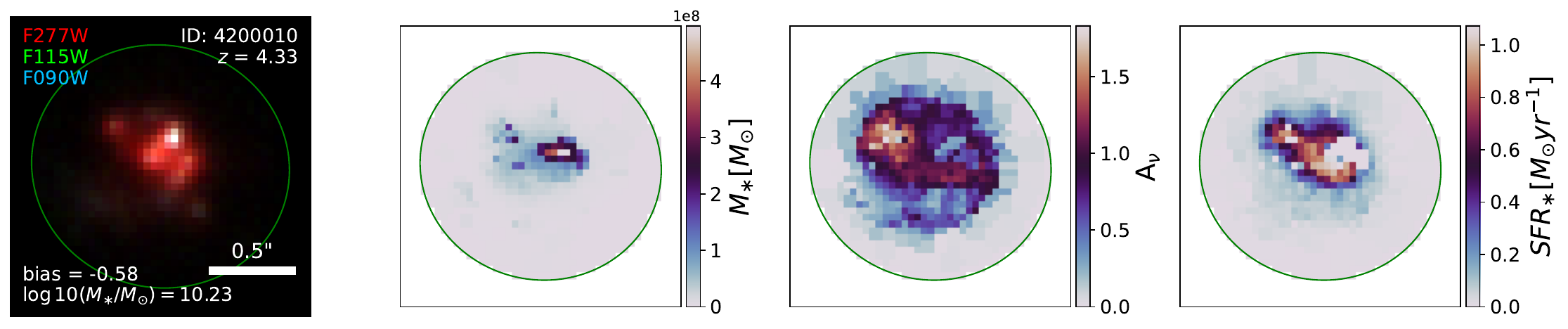}

    \includegraphics[width=\textwidth]{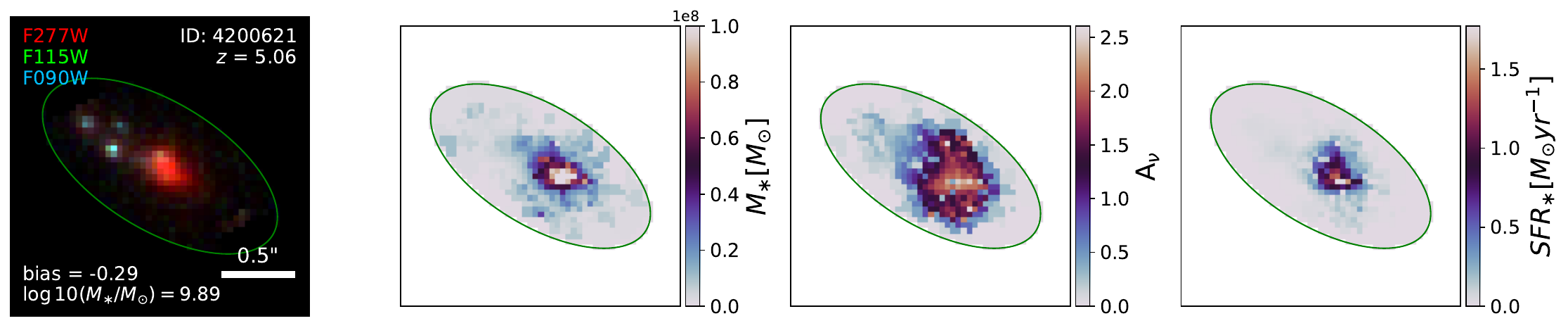}

    \includegraphics[width=\textwidth]{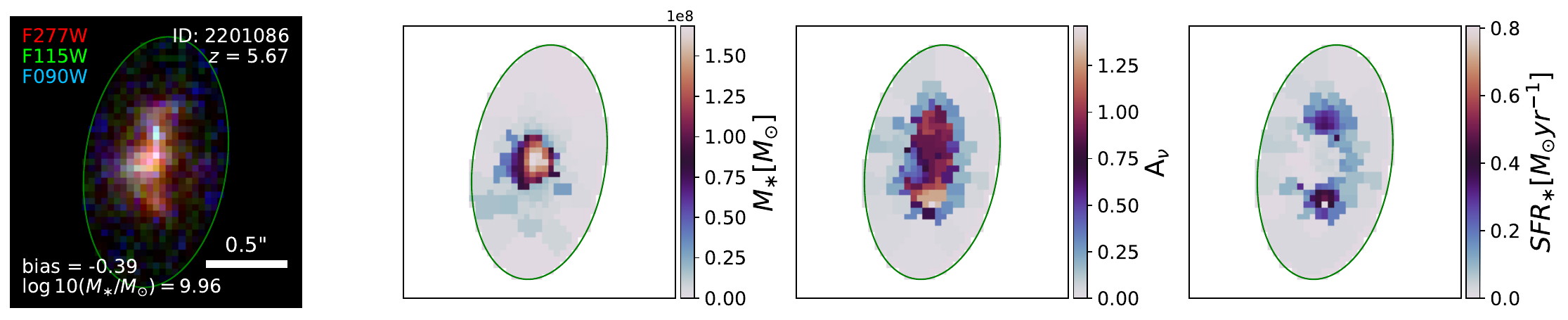}

    \caption{
        Galaxies with a large stellar mass bias. RGB and spatially resolved maps: stellar mass ($M_{\ast}$), dust attenuation ($A_{\nu}$), and star formation rate (SFR$_{\ast}$) of each pixel.}
    \label{fig: rgb and 2d maps large bias}
\end{figure*}

    \begin{figure*}
        \centering
    
        \includegraphics[width=0.49\textwidth]{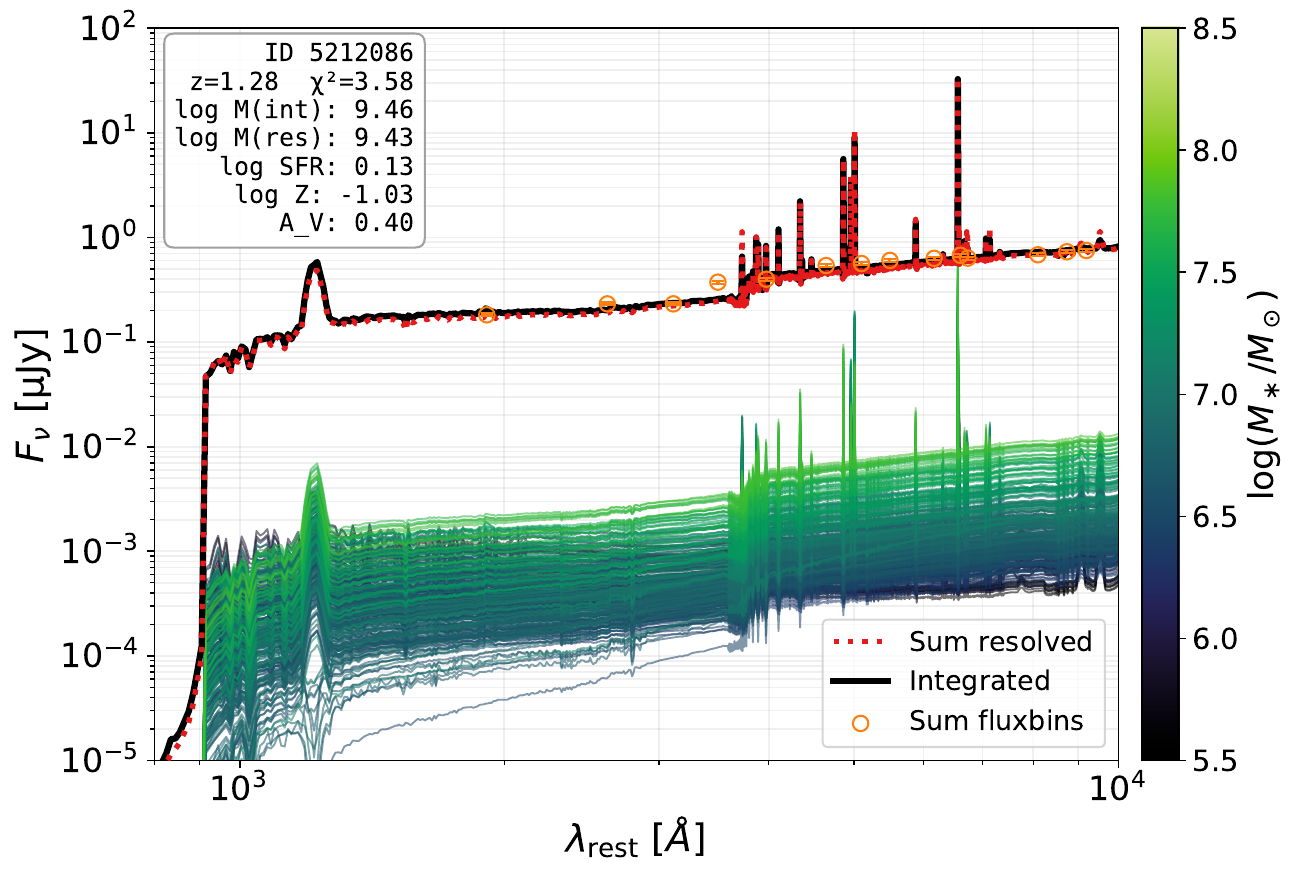}
        \hfill
        \includegraphics[width=0.49\textwidth]{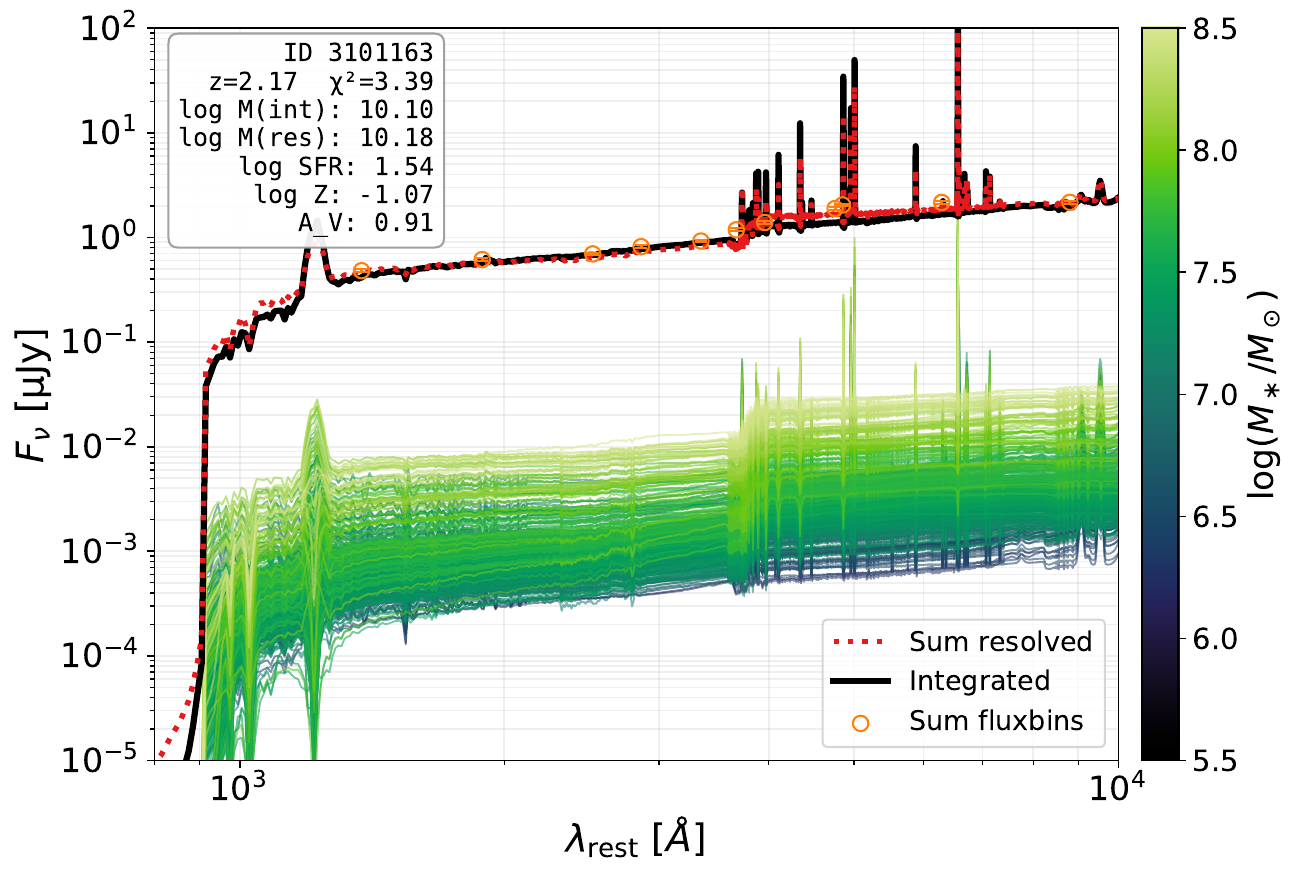}
        \\[0.5em]
        \includegraphics[width=0.49\textwidth]{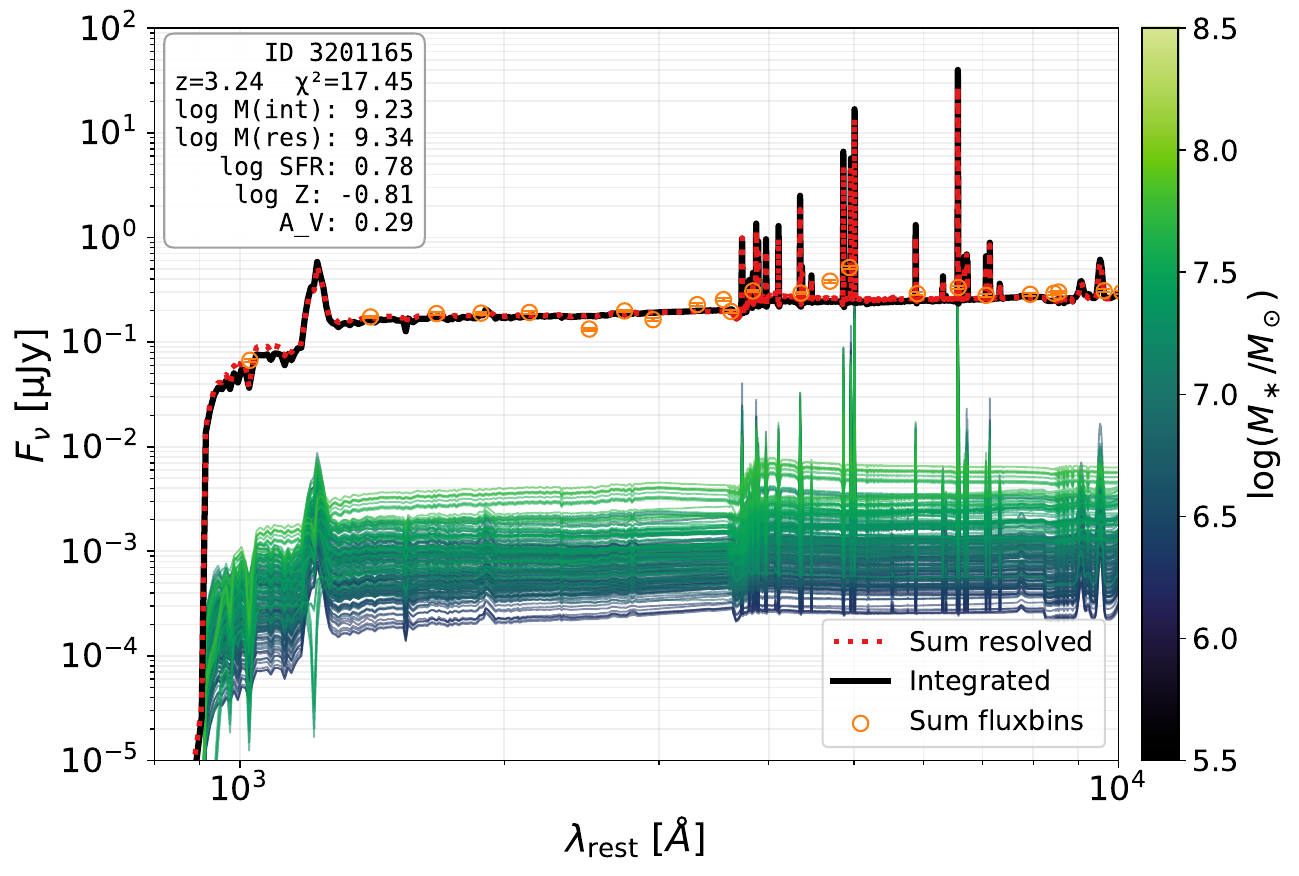}
        \hfill
        \includegraphics[width=0.49\textwidth]{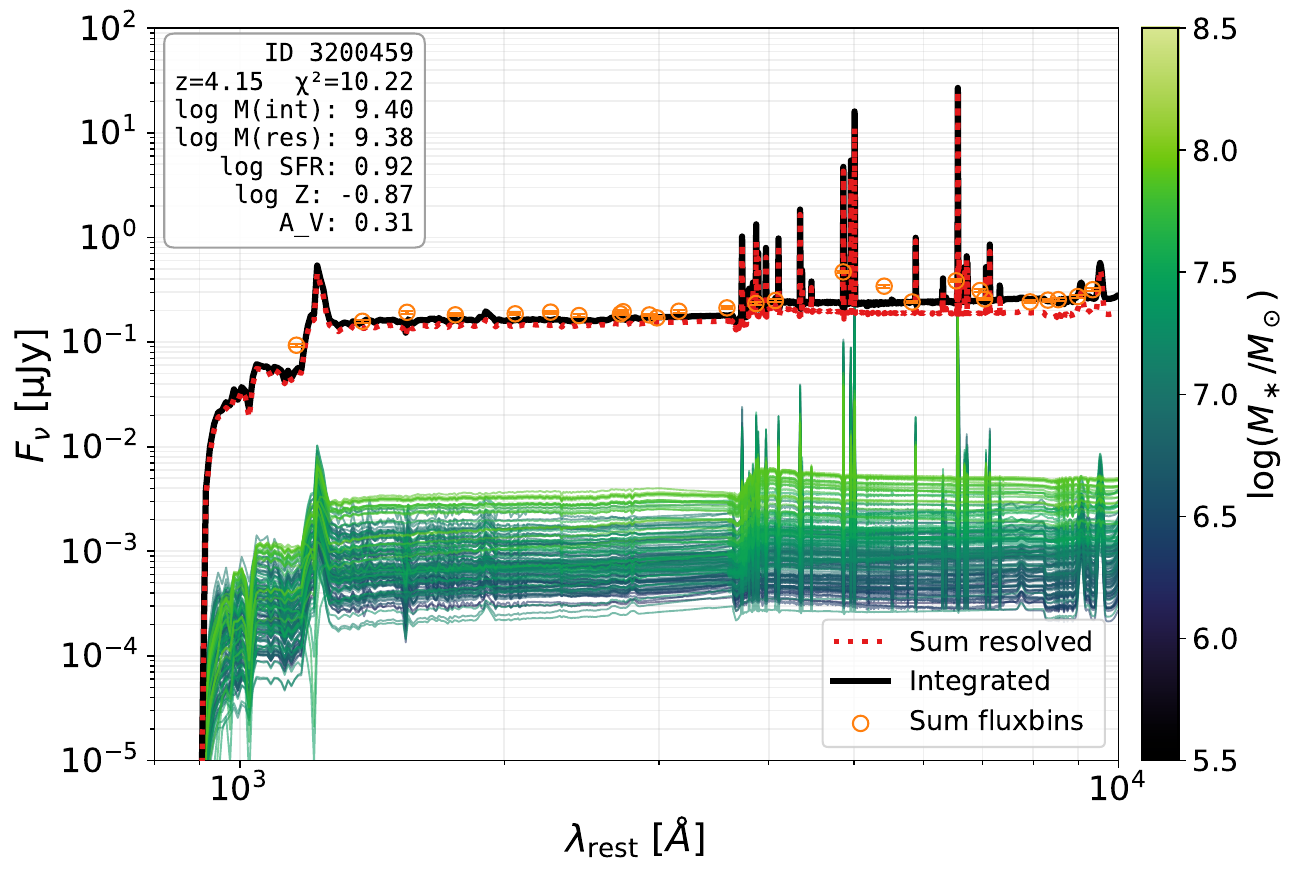}
        \\[0.5em]
        \includegraphics[width=0.49\textwidth]{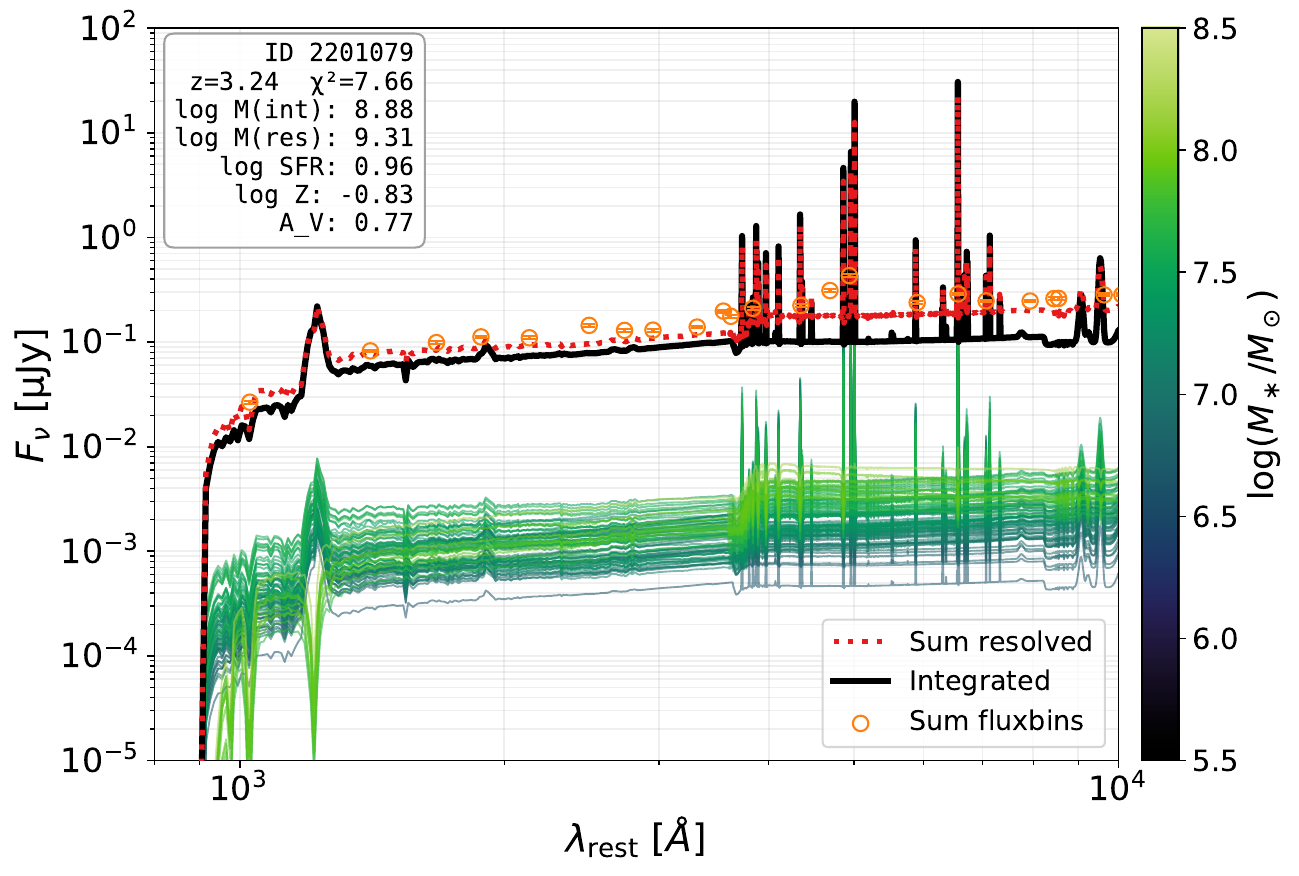}
        \hfill
        \includegraphics[width=0.49\textwidth]{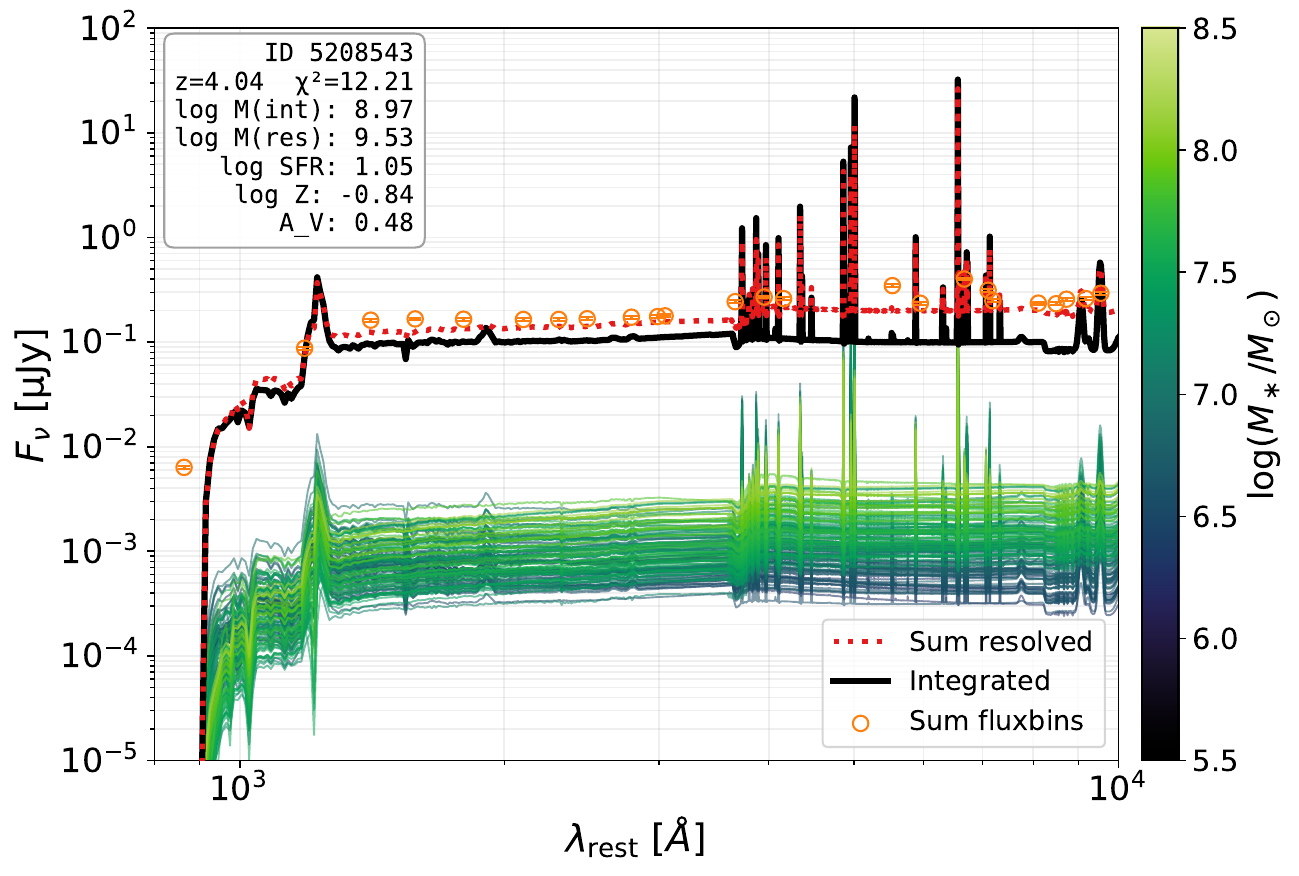}

        \caption{
            Observed spectral energy distributions (SEDs; orange open circles) shown as a function of wavelength. The solid line denotes the best-fitting integrated SED, while coloured curves represent SEDs obtained from spatially resolved Voronoi bins. The dashed line shows the sum of the resolved SEDs as a function of wavelength for six randomly selected galaxies. A low $\chi^2$ can occur despite a poor visual fit because large observational uncertainties reduce the weight of discrepant data points in the $\chi^2$ calculation. Also note that the median posterior stellar parameter values do not necessarily correspond to the parameters of the median posterior spectrum, since each is derived independently from the full posterior distribution.}
    
        \label{fig: int res spec 6 gals}
    \end{figure*}

\subsection{Voronoi binning}\label{subsec: method voronoi}
This work applies the \texttt{Voronoi} binning algorithm of \cite{2003MNRAS.342..345C} to divide each galaxy into spatial bins with approximately uniform signal-to-noise ratio ($S/N_{\mathrm{bin}}$). Our implementation follows a similar method to spatially resolved SED-fitting as \citeauthor{2022ApJ...933...30T} (\citeyear{2022ApJ...933...30T}; \citeyear{2024ApJ...964..177T}) and \cite{2025arXiv250925363S}. Voronoi binning is applied on the F444W JWST/NIRCam image. Tessellations are generated within the Kron aperture of each galaxy, except for systems with effective radii smaller than 0.35", where a fixed circular aperture of 0.35" is used. Contaminating neighbouring sources located inside the aperture are masked, with their pixels replaced by values obtained from corresponding pixels mirrored through the source center to preserve the local background. 

We simplify the pixel-level flux uncertainties used for Voronoi binning by assuming pixels are independent (no noise correlation). Pixel-level flux uncertainties are then estimated from the standard deviation of the background measured from the reversed segmentation map of the science image by fitting a Gaussian, with the extreme tails [0.01, 0.99] trimmed. 
This approach is adopted for computational convenience of the noise, as Voronoi binning requires a S/N threshold to be applied. Although the simplified noise estimate underestimates the true noise because correlated pixels are not accounted for, it is sufficient for constructing the Voronoi tessellation, where the target $S/N_{\mathrm{bin}}$ controls the balance between spatial resolution and photometric precision. Pixel-to-pixel noise correlations are instead accounted for when estimating the uncertainties of the Vorbin fluxes used for SED fitting (see Section \ref{sec: method int res phot}).

The \texttt{Voronoi} algorithm takes as input the flux, flux uncertainty, and ($x$, $y$)-coordinates of each pixel. We adopt a target signal-to-noise ratio of $S/N_{\mathrm{bin}} = 15$ for all bins. Previous resolved photometry studies typically use thresholds of $S/N \approx 5$ (e.g. \citealp{2025MNRAS.542.2998H}; \citealp{2025ApJ...994...94T}), or above. From private communication with K. Iyer, \texttt{Dense Basis} recovers stellar masses with a precision of $\sim$0.2 dex at $S/N_{\mathrm{bin}} \simeq 5$, whereas reliable sSFR constraints require significantly higher quality, around $S/N_{\mathrm{bin}} \simeq 20$ for $\sim$0.3 dex accuracy. Because inferring the stellar mass depends sensitively on resolving the spatial distribution of star formation, we adopt an intermediate threshold of $S/N_{\mathrm{bin}} = 15$ to balance accurate sSFR recovery with sufficient spatial resolution, as our minimum threshold for the total SNR$_{\mathrm{F444W}}$ is 50.

\Cref{fig: voronoi_rgb_segmap_examples} presents two examples of the spatially resolved analysis for galaxies with IDs 2122116 and 4100210. For each galaxy, the figure shows the Voronoi map, the S/N of the individual pixels and resulting Voronoi bins in F444W, the RGB composite image, and the corresponding segmentation map. The S/N panels show the individual pixels in black and the Voronoi bins in red, with unbinned pixels indicated by blue crosses. The dashed blue line marks the target S/N of 15 used by the \texttt{Voronoi} algorithm. In the segmentation maps, the target galaxy is shown in red, while neighbouring target IDs are shown in blue. Neighbouring sources that fall within the Kron aperture are masked and subsequently filled using mirrored pixels. Panels a, b, e, and f are generated by the Voronoi algorithm. \Cref{tab: nr of galaxies} shows that one galaxy was excluded because the Voronoi algorithm could not construct valid bins.

In Appendix~\ref{sec: appendix voronoi and snr threshold}, a comparison between tessellations generated with $S/N_{\mathrm{bin}} = 5$, $S/N_{\mathrm{bin}} = 10$, and $S/N_{\mathrm{bin}} = 15$ shows no significant difference in the stellar mass bins recovered from resolved photometry (see \Cref{fig: log stelm vs snr of vorbins}). This outcome is expected: the integrated stellar mass is dominated by high SNR regions of each galaxy, which inevitably form bins with $S/N_{\mathrm{bin}} > 15$ regardless of the adopted threshold. Lower-$S/N$ pixels have less influence on the summed stellar mass, and are in most cases around the outskirts, at least for galaxies with a non-irregular morphology. Nevertheless, as discussed above, we adopt $S/N_{\mathrm{bin}} = 15$ because it provides more reliable constraints on the stellar mass and sSFR while preserving spatial resolution.

\subsection{SED-fitting with Dense Basis}\label{subsec: method dense basis}
This work uses the \texttt{Dense Basis} SED-fitting code (\citealp{2017ApJ...838..127I}; \citealp{2019ApJ...879..116I}), which differs from traditional parametric frameworks by adopting a non-parametric star-formation history (SFH) defined through stellar mass assembly quantiles in lookback time. We adopt a \citep{2003PASP..115..763C} IMF, \citep{2000ApJ...533..682C} dust attenuation law, and MILES + MIST isochrones and stellar stracks and CLOUDY photoionization models in Flexible Stellar Population Synthesis (FSPS) to model composite stellar and nebular emission (\citealp{2009ApJ...699..486C}; \citealp{2010ApJ...712..833C}). \texttt{Dense Basis} returns several physical parameters, including stellar mass ($M_{\ast}/M_{\odot}$), specific star-formation rate (sSFR), dust attenuation ($A_{\nu}$), metallicity ($Z/Z_{\odot}$), and four lookback times at which different fractions of the stellar mass were assembled ($\alpha_{\mathrm{SFH}}$). For both integrated and resolved fits, we adopt flat priors on all parameters but $\alpha_{\mathrm{SFH}}$, and apply a narrow uniform prior range for the redshift with $\bigtriangleup z$ = 0.01. 
The full set of priors is listed in \Cref{tab: priors atlas dense basis}. An additional 3\% uncertainty is added in quadrature to all photometric flux errors to account for instrumental systematics in JWST and HST imaging.

\begin{table}\centering
\caption[]{\texttt{Dense Basis} priors and parameter choices.}\label{tab: priors atlas dense basis}
\begin{tabular}{ccc} \hline
\textbf{Parameter} & \textbf{Range} & \textbf{Prior} \\ \hline \hline
$\log ({M_{\ast}/M_{\odot}})$ & (5, 12.5) & flat \\ \hline
$\log (\mathrm{sSFR\ [yr^{-1}]})$ & (-14, -5) & flat \\ \hline
$A_V$ & (0, 6) & flat \\ \hline
$\log ({Z/Z_{\odot}})$ & (-2.5, 0.5) & flat \\ \hline
$dz$ & (-0.1, 0.1) & - \\ \hline
$\alpha_{\mathrm{SFH}}$ & 4 & Dirichlet \\ \hline
\end{tabular}
\end{table}

\subsection{Integrated and Resolved Photometry}\label{sec: method int res phot} 
This work compares two photometric methods for determining stellar mass; therefore, a consistent methodology is essential for evaluating differences between integrated and resolved photometry. We apply SED fitting on PSF-convolved images to JWST/NIRCam F444W. Similar to the Kron flux measurements in the DR1 photometry catalog (\citetalias{2026ApJS..282....3S} \citeyear{2026ApJS..282....3S}), we use the flux  measurements within a 2.5 $\times$ Kron aperture (or a fixed circular aperture of 0.35", see Section \ref{subsec: method voronoi}), and correct for Milky Way extinction and PSF aperture effects prior to SED fitting. PSF aperture corrections accounting for flux losses outside the aperture after PSF convolution.

    \begin{figure*}[t]
        \centering
        \includegraphics[width=\textwidth]{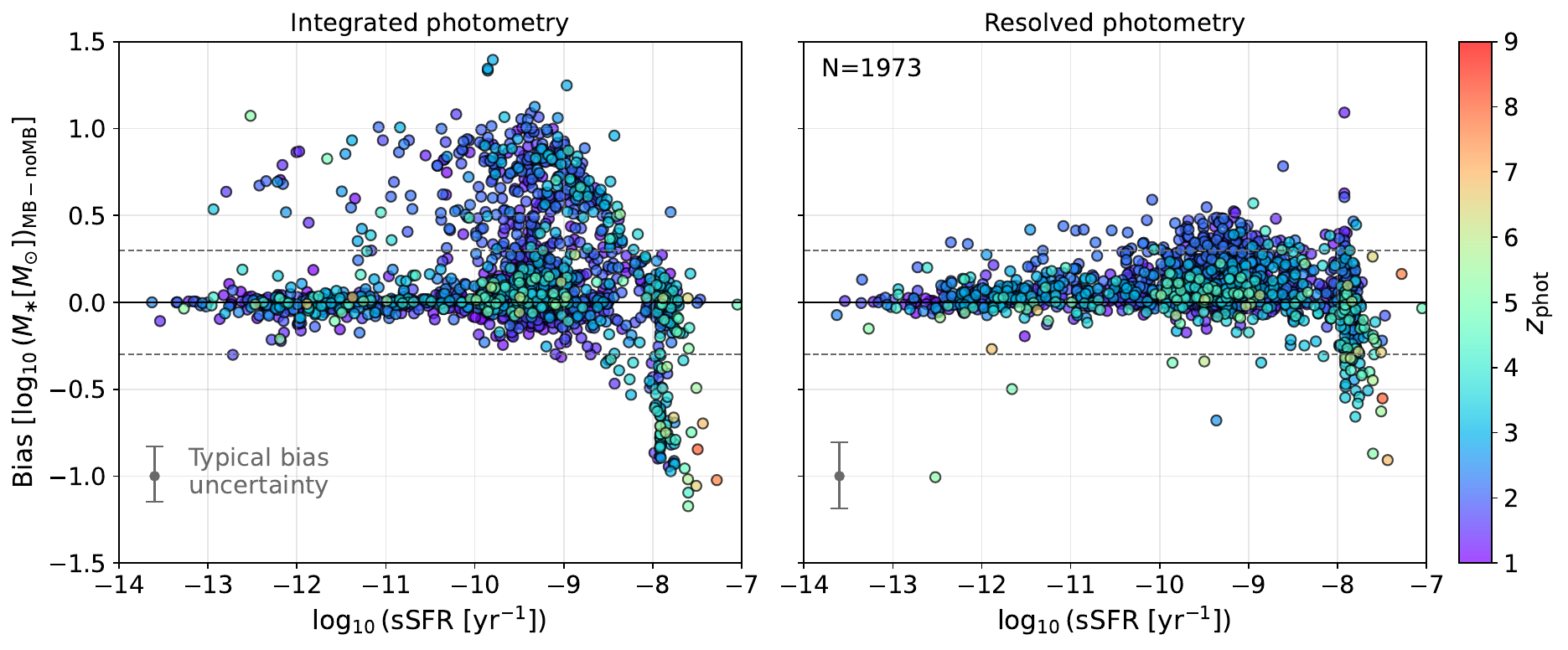}
        \hfill
        \caption{
        Ratio of stellar masses derived with and without medium-band photometry 
        ($M_{\ast,\mathrm{MB}}/M_{\ast,\mathrm{noMB}}$) for integrated (left) and resolved (right) fits, 
        shown as a function of stellar mass surface density. 
        Points are colored by sSFR. 
        Horizontal lines mark unity and the dash lines dex = $\pm$0.3. 
        Resolved photometry yields mass ratios closer to unity and substantially reduced scatter, 
        indicating that spatially resolved modeling mitigates the medium-band–induced biases present 
        in integrated fits. The remaining scatter is larger for smaller galaxies, 
        likely reflecting stronger color gradients, more diverse stellar populations, 
        and increased photometric uncertainties in compact systems. MACS0417-\textit{CLU} and MACS1423-\textit{CLU} are not taking into this sample set, given F410M is the only medium bands for these fields. This leaves the sample set with 1973 galaxies.
        }
        \label{fig: int res ratio medium band}
    \end{figure*}

However, rather than using the DR1 catalog Kron flux directly, we compute integrated photometry as the sum of Voronoi-bin fluxes ("flux bins") within the same apertures. Prior to constructing these bins, the science images are processed in the same manner as for Voronoi binning: adjacent sources within the Kron aperture are masked and replaced with corresponding pixels mirrored through the source center. After applying Voronoi binning, each pixel is assigned to a bin, producing a set of flux bins in each filter, and the total integrated flux is simply given by their sum. The resulting fluxes are then corrected for Milky Way extinction and PSF aperture effects. Thus, stellar masses are derived consistently by applying SED fitting to the summed flux bins for integrated measurements, and by performing SED fitting on individual bins for the resolved case. This method ensures consistency between integrated and resolved stellar mass measurements.

Uncertainties on the flux bins are estimated using a method equivalent to that adopted for single-aperture photometry in the DR1 catalog (\citetalias{2026ApJS..282....3S} \citeyear{2026ApJS..282....3S}), in order to account for pixel-to-pixel correlations introduced by the drizzling process. In drizzled JWST images, noise is correlated due to image resampling and stacking, such that photometric uncertainties cannot be reliably obtained by simply summing pixel variances under the assumption of independent pixels. 

Instead, flux uncertainties are derived from empirical background measurements in DR1. Briefly, 2000 empty apertures are placed on background regions of the science image using fixed diameters of 0.3", 0.5", 0.7", 1.0", and 1.5", and the standard deviation of the resulting flux distributions is measured via Gaussian fitting. These measurements are then scaled to match the effective areas of the Voronoi bins and the full Kron aperture, thereby providing uncertainty estimates for both resolved (per-bin) and integrated fluxes.

With these fluxes and associated uncertainties, integrated and resolved stellar masses are then derived by fitting the spectral energy distributions (SEDs) using \texttt{Dense Basis}, incorporating both the photometric uncertainties and an additional 3\% systematic uncertainty. \Cref{fig: flowchart method} in Appendix~\ref{sec: appendix flowchart method} demonstrates a flowchart of all the steps taken from science image to integrated and resolved photometry. \Cref{fig: rgb and 2d maps small bias} and \Cref{fig: rgb and 2d maps large bias} illustrate examples of the resulting two-dimensional parameter maps for a subset of galaxies in the sample. For each galaxy, the panels display the stellar mass $M_{\ast}$, dust attenuation $A_{\nu}$, and SFR. These systems are relatively dusty and exhibit spatial variations in both mass and star-formation.

\Cref{fig: int res spec 6 gals} shows the SEDs generated by \texttt{Dense Basis} for six randomly selected galaxies, four of which show no significant bias in stellar mass and two of which show a negative bias. For each galaxy, the integrated SED is compared with the sum of the  resolved SEDs, the resolved SED bins are also demonstrated. The data points represent the sum of the fluxes in the individual resolved flux bins, allowing the resolved measurements to be compared directly with the integrated fit. In the two galaxies shown in the lower panels, the resolved stellar mass is larger than the integrated stellar mass, resulting in the resolved SED lying above the integrated SED. In these cases, the summed resolved flux bins are also better reproduced by the resolved SED than by the integrated SED. The four galaxies without a significant stellar mass bias, shown in the upper panels, show better agreement between the integrated and resolved results. Note that a low $\chi^2$ can occur despite a poor visual fit because large observational uncertainties reduce the weight of discrepant data points in the $\chi^2$ calculation. The median posterior stellar parameters may differ from those associated with the median posterior spectrum, as they represent different summaries of the posterior distribution. Magnification from gravitational lensing in the \textit{CLU} fields has been applied after photometry to the stellar parameters where appropriate ($M_{\ast}$, luminosity, size, and SFR).
    
\section{Results}\label{sec: results}

\subsection{Stellar Mass Estimates and Medium Bands}\label{subsec: stellar mass and medium bands}

Before studying the bias between integrated and resolved stellar mass estimates, 
we compare results obtained from wide-band photometry alone with those including medium-band filters to quantify the effect of medium-band coverage on the inferred stellar masses. This analysis uses HST ACS/WFC and JWST NIRCam imaging of the five CANUCS fields and their parallel fields. The available medium-band filters vary by field (see \Cref{tab:clusters_and_filters}), and include F140M, F162M, F182M, F210M, F250M, F300M, F335M, F410M, F430M, F460M, and F480M. Given the lack of medium-bands (only F410M), the fields MACS0417-\textit{CLU} and MACS1423-\textit{CLU} are not included, which leaves the sample set with 1942 galaxies.

\Cref{fig: int res ratio medium band} shows the difference in stellar mass estimates between results with and without medium bands for both integrated (left panel) and resolved (right panel) photometry as a function of sSFR. The colorbar indicates $z_{\mathrm{phot}}$ of each galaxy measured with medium band inclusion. We find that spatially resolved photometry reduces the offset in stellar mass estimates obtained with and without medium bands compared to integrated photometry. The standard deviation of the bias distribution is $\sigma_{\mathrm{int}} = 0.33$ dex for integrated photometry and $\sigma_{\mathrm{res}} = 0.15$ dex for spatially resolved photometry. We find no clear correlation between the bias and galaxy compactness. Instead, the largest overestimates in stellar mass when medium bands are omitted are mainly associated with galaxies with intermediate to high specific star-formation rates, $\log_{10}(\mathrm{sSFR}/\mathrm{yr}^{-1}) \lesssim -8$.

\cite{2024ApJ...967L..17S} reported that excluding higher spectral resolution medium band filters from integrated photometry in MACS0417 \textit{NCF} leads to an overestimation of the inferred stellar mass density of galaxies at high redshift ($z > 5$). Similarly, \cite{2025MNRAS.542.2998H} tested integrated photometry for $\sim$200 galaxies in the JADES field and observed comparable trends. \cite{2024ApJ...967L..17S} further discussed that including medium-bands into photometry improves SED-fitting by distinguishing emission lines from the continuum at specific redshifts, thereby reducing the overestimation of stellar masses. In other words, \Cref{fig: int res ratio medium band} shows that performing photometry on spatially resolved regions of a galaxy seems to prevent localized regions with strong emission lines from inflating the total stellar mass, an effect that can arise when using broadband photometry alone.

\subsection{Stellar Mass Bias and Galaxy Properties}\label{subsec: stellar mass bias int res}

    \begin{figure*}
        \centering
        \includegraphics[width=\textwidth]{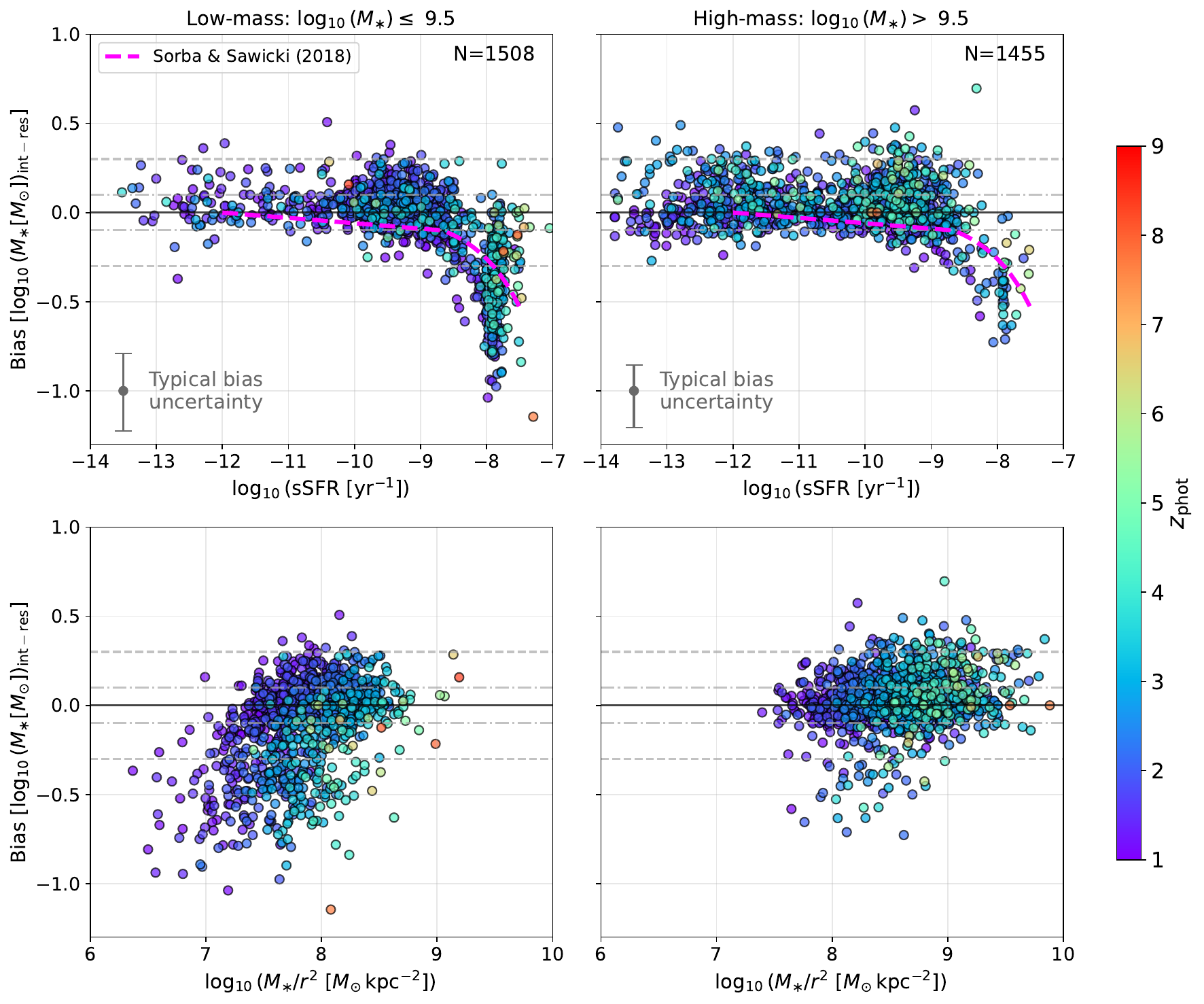}
        \hfill
        \caption{
        Stellar mass bias ($\log_{\mathrm{10}} M_{\mathrm{int}} - \log_{\mathrm{10}} M_{\mathrm{res}}$) for galaxies with medium-band photometry ($\mathrm{MB}=\mathrm{True}$), split by stellar mass. 
        The top panels show the stellar mass bias versus $\log_{10}(\mathrm{sSFR})$ for low-mass ($\log_{10} M_\ast \leq 9.5$) and high-mass ($\log_{10} M_\ast > 9.5$) galaxies. 
        The bottom panels show the bias versus stellar mass surface density ($\log_{\mathrm{10}}(M_\ast / R^2)$), highlighting lack of resolution in high density systems. 
        Points are color-coded by photometric redshift. 
        Horizontal lines indicate zero bias and $\pm0.1$, $\pm0.3$~dex reference levels. 
        Overall, low-mass galaxies exhibit larger mass biases and lower $M_\ast / R^2$, while high-mass systems show smaller scatter, suggesting that morphology and spatial distribution of stellar mass contribute to differences between integrated and resolved photometry. The dashed line is from \cite{2018MNRAS.476.1532S}, please note that there is no distinction between mass bins for this line.
        }\label{fig: bias sSFR mass size zphot}
    \end{figure*}

We investigate systematic differences in stellar mass estimates derived from integrated and spatially resolved photometry, with both measurements estimated using \texttt{Dense Basis} (Section \ref{sec: method int res phot}). Understanding these differences is important for assessing potential biases in stellar mass measurements across diverse galaxy populations. \Cref{fig: bias sSFR mass size zphot} shows the stellar mass bias, defined as the difference between the logarithmic stellar masses obtained from integrated and resolved photometry - bias = $\log10(M_{\ast,\mathrm{int}}) - \log10(M_{\ast,\mathrm{res}})$, as a function of specific star-formation rate (sSFR) and stellar mass surface density ($M_{\ast}/r^{2}$), with data points color-coded by photometric redshift. The quoted bias uncertainties correspond to the central 68\% interval, estimated from the 16th and 84th percentile uncertainties of the resolved and integrated stellar-mass estimates. The dashed line is from \cite{2018MNRAS.476.1532S}, please note that there is no distinction between mass bins for this line. 

The sample is divided at $\log_{10}(M_{\ast}/M_{\odot}) = 9.5$, the median of the stellar mass bias of $\log_{10}(M_{\ast}/M_{\odot}) < 9.5$ is -0.03 dex and 0.03 dex of $\log_{10}(M_{\ast}/M_{\odot}) > 9.5$. The difference between the mass bins is small due to being compressed by the full distribution. However, the lower mass bin - which has more galaxies with a large negative bias - contains more galaxies with a higher sSFR. 
\Cref{tab: median bias ssfr bin} lists the median values of the stellar mass bias in sSFR bins. The median stellar mass bias decreases from $-0.384$ dex in the highest sSFR bin to approximately zero in the intermediate bin, before becoming slightly positive at lower sSFRs ($0.025$ and $0.033$ dex). Thus, the typical integrated stellar mass is substantially underestimated for the highest-sSFR galaxies, whereas the median bias is negligible for galaxies with lower sSFRs.

Previous studies have reported that differences between stellar masses derived from integrated and spatially resolved photometry correlate with star-formation activity (\citeauthor{2018MNRAS.476.1532S} \citeyear{2018MNRAS.476.1532S}; \citeauthor{2023ApJ...948..126G} \citeyear{2023ApJ...948..126G}; \citeauthor{2025MNRAS.542.2998H} \citeyear{2025MNRAS.542.2998H}). These works connect the relation primarily to the presence of bright, young stellar populations that dominate the integrated light in actively star-forming systems. In \Cref{fig: bias sSFR mass size zphot} we recover a similar correlation between the stellar mass offset and sSFR. However, the dependence on photometric redshift appears weaker in our analysis than reported by \cite{2018MNRAS.476.1532S}.

The lower panels in \Cref{fig: bias sSFR mass size zphot} reveal a correlation between the bias and stellar mass surface density. Low-density galaxies show larger biases, whereas high-density systems exhibit smaller biases. As galaxies become more compact it becomes more difficult to resolve separate stellar populations, and therefore resolved and unresolved masses become asymptotically similar. This does not necessarily mean the unresolved mass is a good description of the total mass, indeed it may just reflect that both resolved and unresolved become similarly incorrect.

    \begin{table}[]
        \centering
        \caption{Median stellar mass bias for each sSFR bin.}\label{tab: median_bias_ssfr}
        \begin{tabular}{cc}
            \hline
            $\log_{10}(\mathrm{sSFR})$ range & Median stellar mass bias [dex] \\
            \hline \hline
            $-8 < \log_{10}(\mathrm{sSFR}) < -7$   & -0.384 \\
            $-9 < \log_{10}(\mathrm{sSFR}) < -8$   & -0.007 \\
            $-10 < \log_{10}(\mathrm{sSFR}) < -9$  & 0.025 \\
            $-14 < \log_{10}(\mathrm{sSFR}) < -10$ & 0.033 \\
            \hline
        \end{tabular}\label{tab: median bias ssfr bin}
    \end{table}

\subsection{2D Mapping: Mass-to-Light Variations}\label{subsec: mass-to-light variations}
\begin{figure*}
    \centering

    \includegraphics[width=\textwidth]{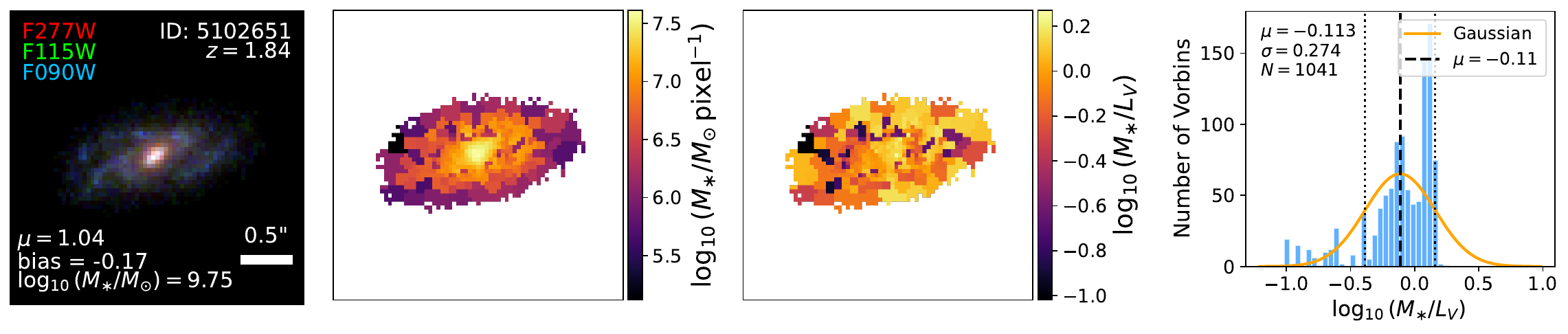}

    \includegraphics[width=\textwidth]{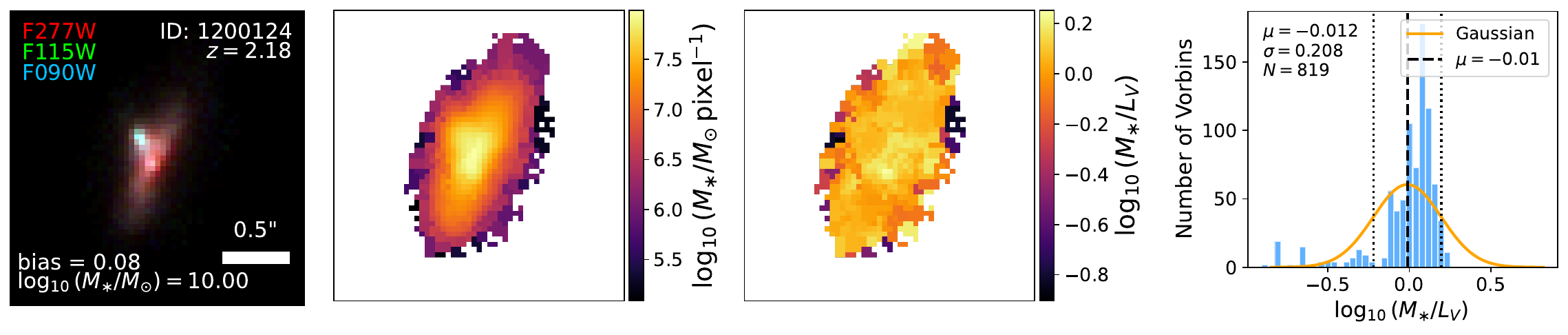}

    \includegraphics[width=\textwidth]{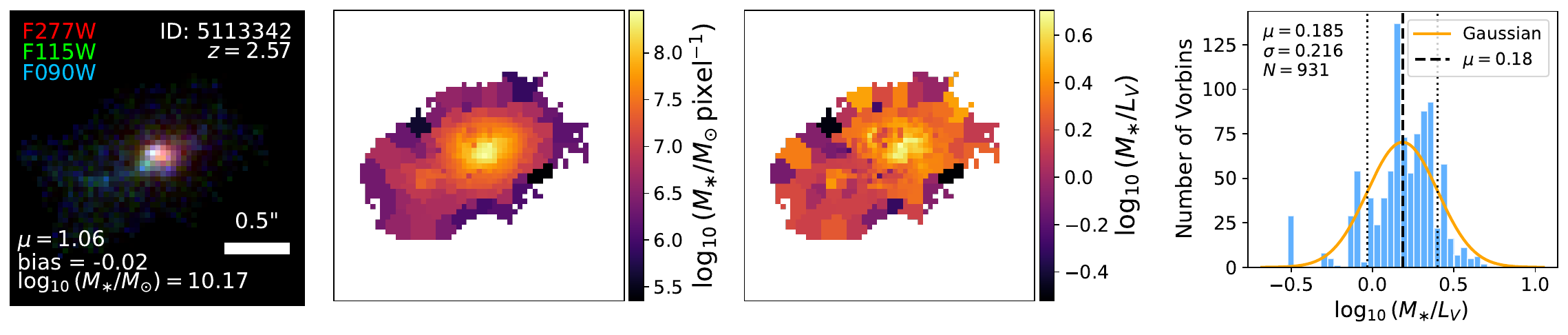}

    \includegraphics[width=\textwidth]{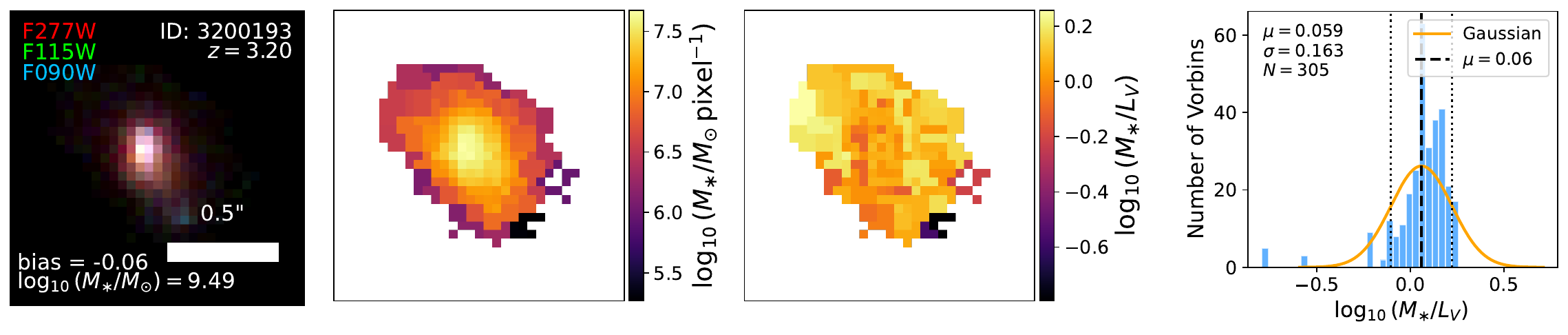}

    \includegraphics[width=\textwidth]{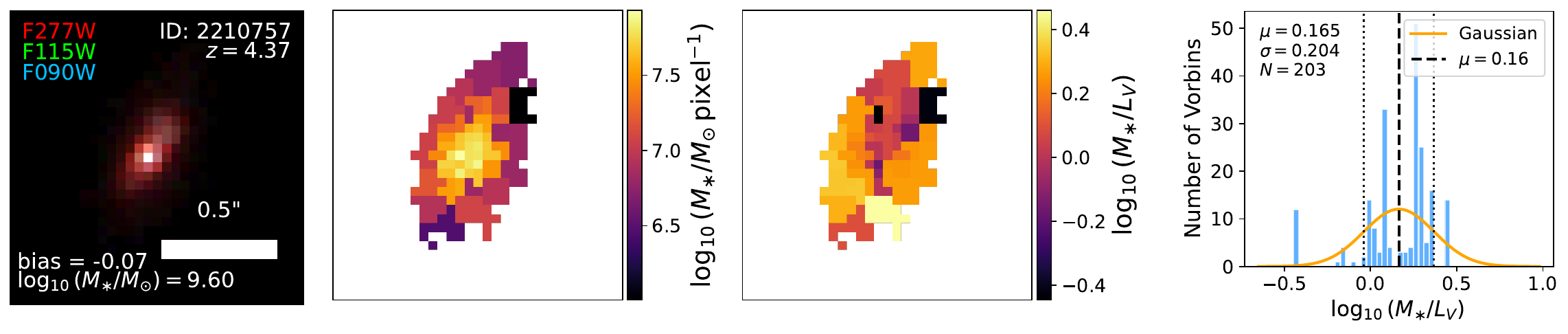}

    \includegraphics[width=\textwidth]{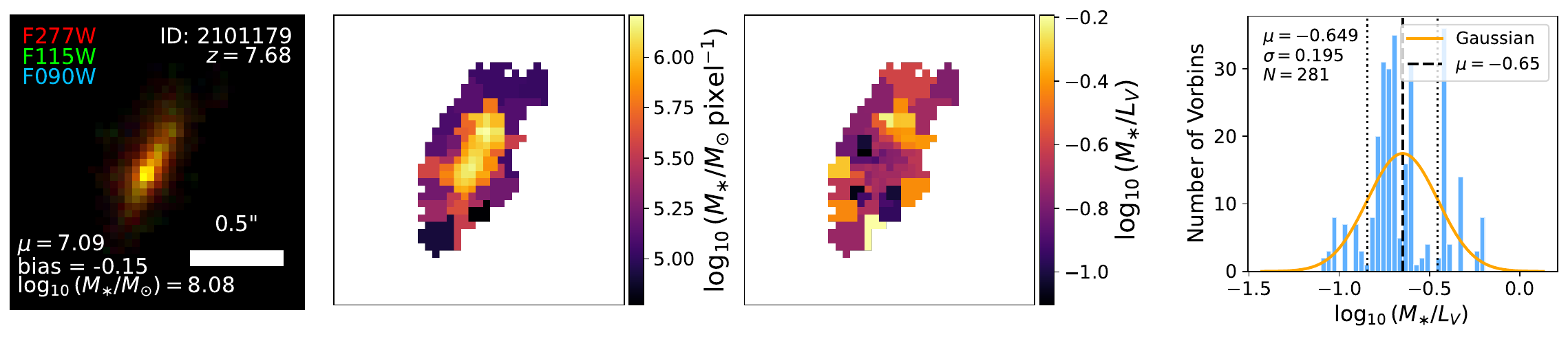}

    \caption{
        Galaxies with a small stellar mass bias. RGB segmentation and spatially resolved maps---stellar mass $\log_{\mathrm{10}}(M_{\ast}/M_{\odot})$ of each pixel and the mass-to-light ratio $\log_{\mathrm{10}}(M_{\ast}/L_{V})$ of each bin---and the histogram of $\log_{\mathrm{10}}(M_{\ast}/L_{V})$. Same galaxies as \Cref{fig: rgb and 2d maps small bias}.
    }
    \label{fig: rgb hist and 2d maps small bias}
\end{figure*}

\begin{figure*}
    \centering

    \includegraphics[width=\textwidth]{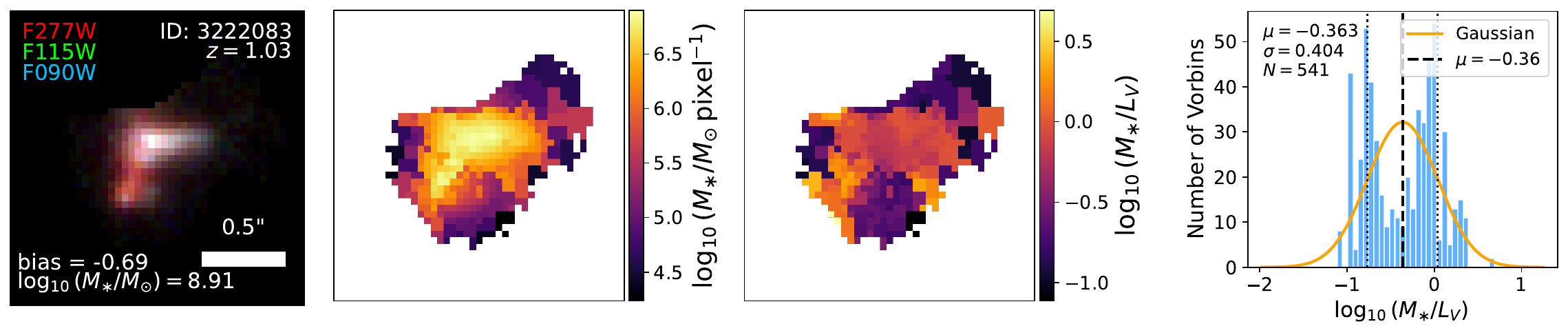}

    \includegraphics[width=\textwidth]{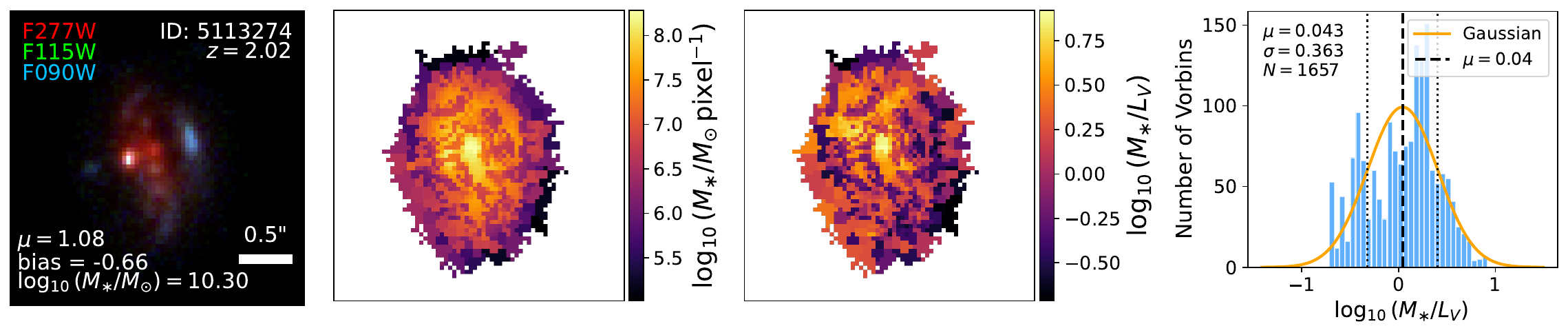}

    \includegraphics[width=\textwidth]{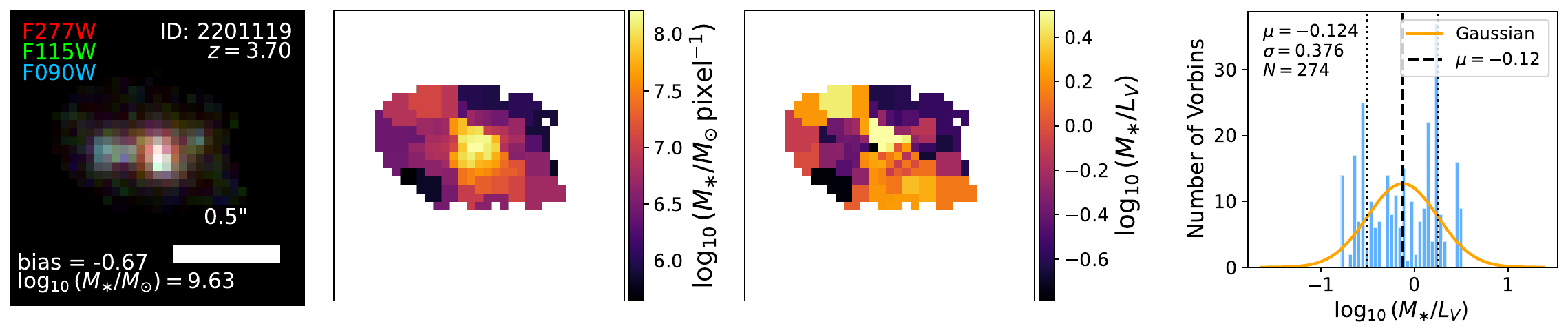}

    \includegraphics[width=\textwidth]{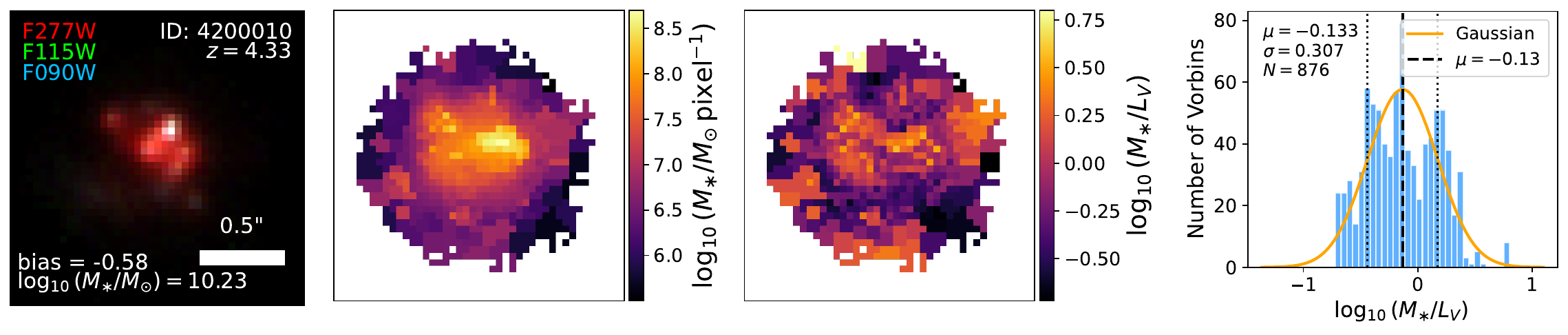}

    \includegraphics[width=\textwidth]{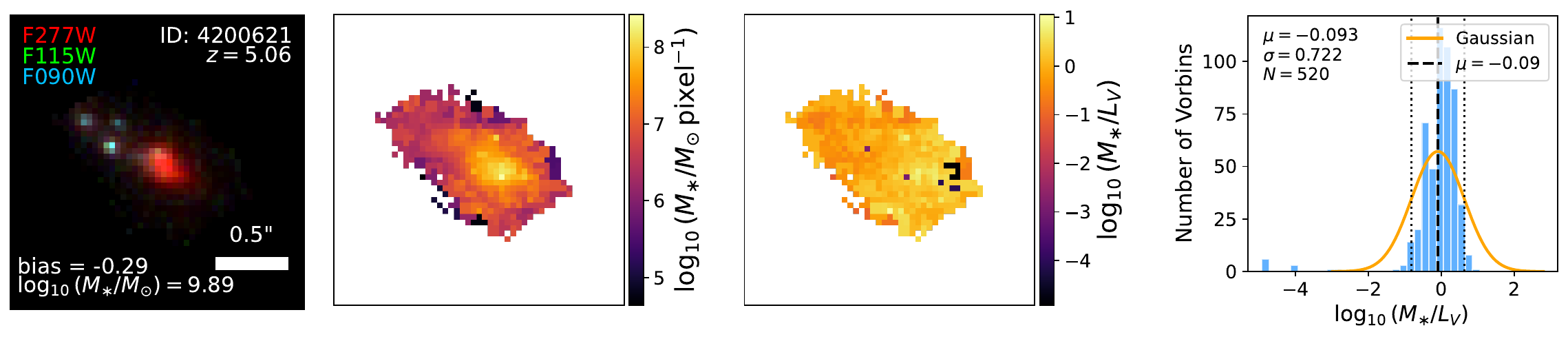}

    \includegraphics[width=\textwidth]{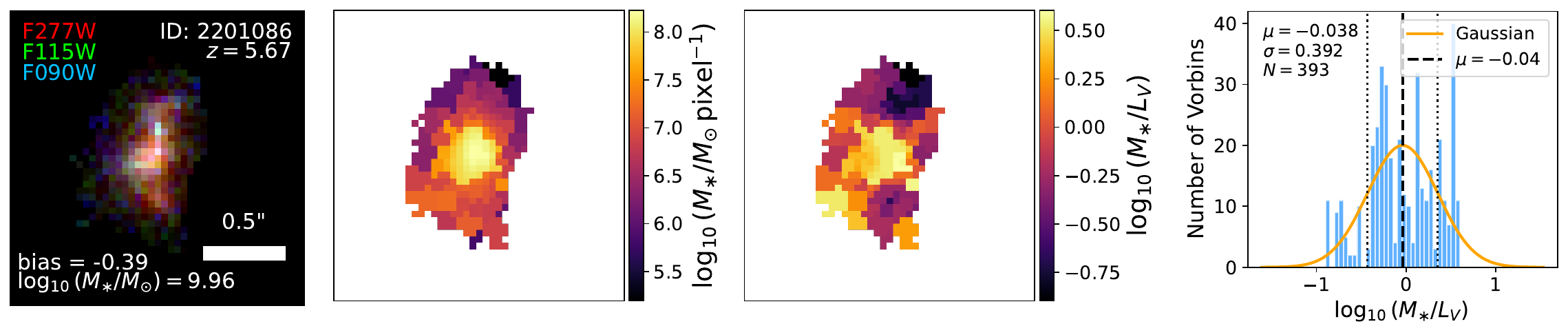}

    \caption{
        Galaxies with a large stellar mass bias. RGB segmentation and spatially resolved maps---stellar mass $\log_{\mathrm{10}}(M_{\ast}/M_{\odot})$ of each pixel and the mass-to-light ratio $\log_{\mathrm{10}}(M_{\ast}/L_{V})$ of each bin---and the histogram of $\log_{\mathrm{10}}(M_{\ast}/L_{V})$. Same galaxies as \Cref{fig: rgb and 2d maps large bias}.
    }
    \label{fig: rgb hist and 2d maps large bias}
\end{figure*}

\begin{figure*}
    \centering
    \includegraphics[width=\textwidth]{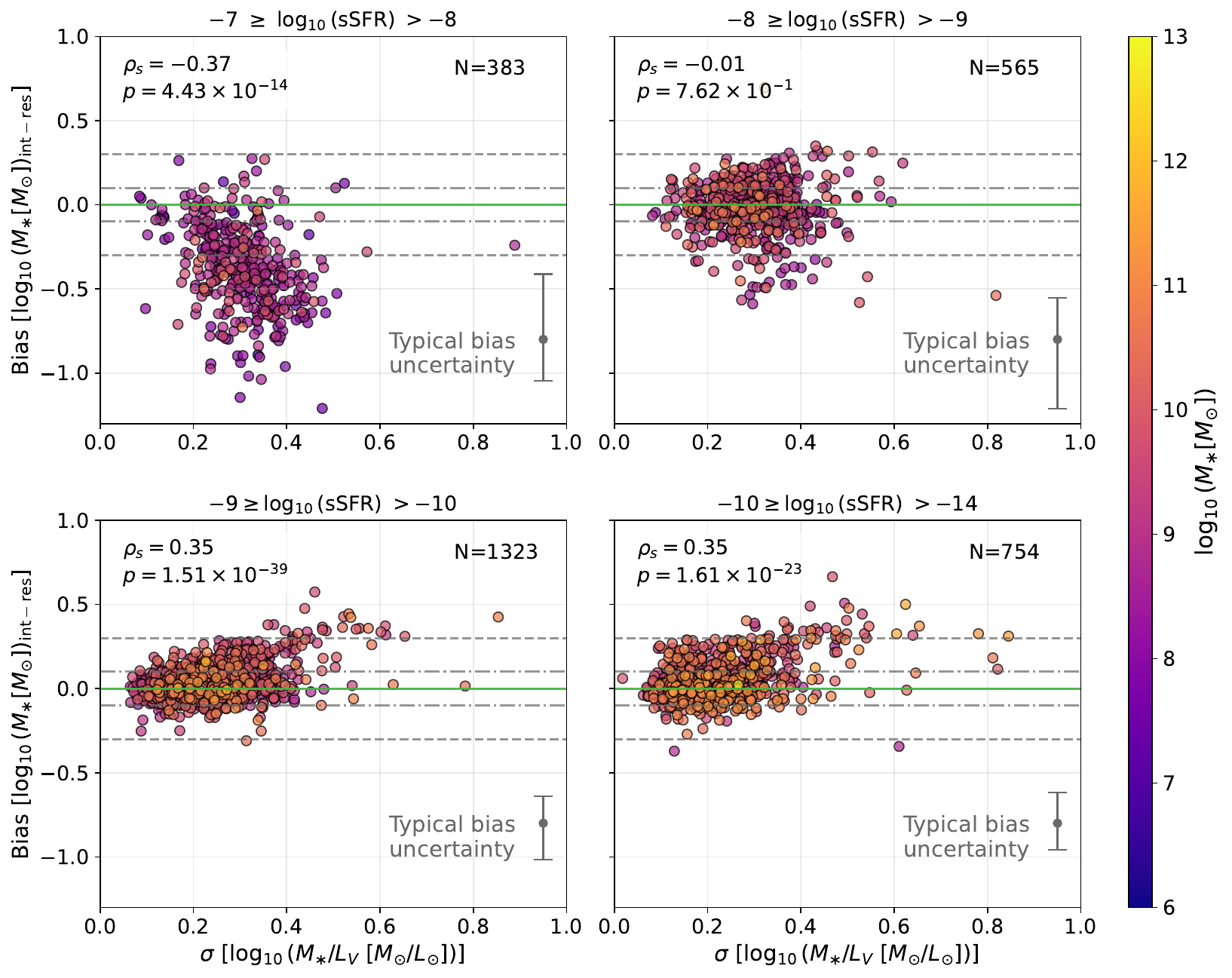}
    \hfill
    \caption{Stellar mass bias, defined as $\log M_{\mathrm{int}} - \log M_{\mathrm{res}}$, shown as a function of the standard deviation of the spatially resolved $\log (M_\ast/L_V)$ distribution. Points are colour-coded by $\log_{10}(\mathrm{M_{\odot}}\,[M_{\ast}])$, as indicated by the color bar. Horizontal dashed lines mark zero bias and reference levels at $\pm 0.1$ and $\pm 0.3$ dex. Galaxies with larger standard deviation at high sSFR (10$^\mathrm{-8}$ yr$^{-1}$), corresponding to more asymmetric internal distributions, tend to exhibit larger stellar mass biases.}
    \label{fig: bias stellar mass vs std mass-to-light ratios}
\end{figure*}

While previous studies have established correlations between the outshining-driven stellar mass bias and global galaxy properties such as sSFR and star formation history, the spatial origin of this effect within galaxies remains less well constrained. \cite{2018MNRAS.476.1532S} reported that the stellar mass bias increases with sSFR, a trend later confirmed by \cite{2025MNRAS.542.2998H}. \cite{2025MNRAS.542.2998H} further showed that bursty star-formation histories, particularly recent bursts, are strongly correlated with the stellar mass bias.

\Cref{fig: rgb and 2d maps small bias} (galaxies with small bias) and \Cref{fig: rgb and 2d maps large bias} (galaxies with large bias) present two-dimensional distributions of stellar mass $M_{\ast}$, dust $A_{V}$, and star formation SFR$_{\ast}$ for our sample, illustrating the intrinsically non-uniform spatial structures of high-redshift galaxies. These maps highlight the coexistence of diverse stellar populations across different regions. While previous studies have linked the outshining effect to global galaxy properties, its spatial origin within galaxies - specifically where it arises and why certain regions dominate the integrated light - remains poorly constrained. Rather than relying solely on star formation histories inferred from SED fitting, this work uses direct spatially resolved 2D observations to investigate the internal origin of the effect. In doing so, we reveal spatial variations in the mass-to-light ($M_{\ast} / L_{V}$) ratio across individual galaxies, which represent the key observable manifestation of outshining.

To quantify these variations, the V-band flux is computed by convolving the redshifted spectral energy distribution (SED) with the filter transmission curve, as given by \Cref{eq: V band flux transmission}. The input SED corresponds to the best-fitting model obtained from the \texttt{Dense Basis} SED-fitting procedure, ensuring consistency with the stellar masses derived using the same method. The \texttt{Dense Basis} SED is provided in $\mu$Jy and is converted to wavelength-based flux. For the V-band, we adopt the Johnson filter transmission curve as implemented in \texttt{EAzY}. The resulting quantity, $F_{\,\mathrm{V}}$, represents the flux integrated over the V-band filter. The corresponding luminosity is then obtained from the luminosity distance via \Cref{eq: luminosity V band from flux},
\begin{equation}\label{eq: V band flux transmission}
F_{\,\mathrm{V}} =
\frac{\int F^{\mathrm{obs}}_{\lambda}\!(\lambda)\, T_{\mathrm{V}}(\lambda)\, d\lambda}
{\int T_{\mathrm{V}}(\lambda)\, d\lambda}
\end{equation}\\

\begin{equation}\label{eq: luminosity V band from flux}
L_{\,\mathrm{V}} =
4\pi D_L^2\, F_{\mathrm{V}},
\end{equation}

where $D_{\mathrm{L}}$ is the luminosity distance. The stellar mass-to-light ratio in the V-band, $M_{\ast}/L_{\mathrm{\nu,V}}$ (hereafter: $M_{\ast}/L_{\mathrm{V}}$), is then obtained by combining the resolved stellar mass obtained from \texttt{Dense Basis} with the corresponding V-band luminosity derived from the bin-wise fluxes following \Cref{eq: V band flux transmission} and \Cref{eq: luminosity V band from flux}, ensuring that both quantities are consistently defined within the same modelling framework. 

Having defined the stellar mass-to-light ratio, we now examine its spatial distribution within galaxies. \Cref{fig: rgb hist and 2d maps small bias} and \Cref{fig: rgb hist and 2d maps large bias} show the RGB segmentation, spatially resolved stellar mass and mass-to-light ratio maps, and the corresponding distributions of $\log_{10}(M/L_V)$ for the same galaxies in \Cref{fig: rgb and 2d maps small bias} and \Cref{fig: rgb and 2d maps large bias}, respectively. The spatially resolved maps show the stellar mass, $\log_{10}(M_{\ast}/M_{\odot})$, for each pixel and the mass-to-light ratio, $\log_{10}(M/L_V)$, for each spatial bin. The histogram shows the distribution of $\log_{10}(M/L_V)$ values across the spatial bins. Note that $\sigma_{\log_{\mathrm{10}}(M_{\ast}/L_{V})}$ is derived within the segmentation of the galaxy to avoid bins in the outskirts, with a low stellar mass and mainly consisting of background, to skew the distribution.

\Cref{fig: bias stellar mass vs std mass-to-light ratios} shows the stellar mass bias as a function of the standard deviation of $\log_{\mathrm{10}}(M_{\ast}/L_V)$ for galaxies separated into four sSFR bins, with the colorbar displaying the total stellar mass. The highest sSFR bin ($-7 < \log_{\mathrm{10}}(\mathrm{sSFR}) < -8$) exhibits a correlation, with galaxies showing a larger dispersion in $\log_{\mathrm{10}}(M_{\ast}/L_V)$ also displaying a larger stellar mass bias. A weaker trend is present for $-8 < \log_{\mathrm{10}}(\mathrm{sSFR}) < -9$, while no correlation is observed in the lower sSFR bins ($-9 < \log_{\mathrm{10}}(\mathrm{sSFR}) < -10$ and $-10 < \log_{\mathrm{10}}(\mathrm{sSFR}) < -14$). The corresponding Spearman rank coefficients are $\rho_{\rm S}=-0.37$, $-0.01$, $0.35$, and $0.35$, respectively, indicating a moderate negative correlation in the highest sSFR bin, negligible correlation in the second bin, and moderate positive correlations in the two lowest sSFR bins.

Galaxies with higher sSFR contain younger stellar populations and consequently exhibit greater spatial variations in their stellar populations, resulting in a broader distribution of mass-to-light ratios. The galaxies in the highest sSFR bins are also systematically lower in stellar mass than those in the lower sSFR bins. The increasing stellar mass bias with increasing $\log_{\mathrm{10}}(M_{\ast}/L_V)$ dispersion in these galaxies is consistent with the outshining effect, in which young, luminous stellar populations dominate the integrated light and lead to an underestimation of the total stellar mass.

\section{Discussion}\label{sec: discussion}

\subsection{Systematic Biases in Integrated Stellar Mass Estimates}\label{subsec: morphology outhshining effect}

The comparison between integrated and spatially resolved photometry provides a useful way to assess potential systematics in stellar mass estimates derived from integrated galaxy measurements, as shown in \Cref{fig: bias sSFR mass size zphot}, particularly for star-forming galaxies. Consistent with previous studies (\citeauthor{2018MNRAS.476.1532S} \citeyear{2018MNRAS.476.1532S}; \citeauthor{2023ApJ...948..126G} \citeyear{2023ApJ...948..126G}) and the recent analysis by \citet{2025MNRAS.542.2998H}, galaxies with high sSFR and bursty or spatially concentrated star formation tend to show larger discrepancies, as young, luminous populations dominate the flux. 

These systematic biases may also have implications for measurements of the cosmic stellar mass density, which is commonly derived from stellar mass functions based on integrated stellar mass estimates. Because the biases identified here depend on galaxy properties such as sSFR and $M_{\ast}$, they may affect different populations non-uniformly, potentially altering both the shape and normalization of the stellar mass function. This could in turn influence comparisons between the stellar mass density inferred from mass functions and that expected from the integrated cosmic star formation history. Quantifying this effect is beyond the scope of this work, but represents an interesting direction for future study.

\texttt{Dense Basis} SED fitting captures the outshining effect consistently across our sample. Integrated measurements can therefore be more strongly affected by the presence of bright, young stellar populations, whose low mass-to-light ratios may bias the inferred stellar mass when different stellar populations are blended within a single aperture.
Spatially resolved photometry mitigates this effect- to some extend- by allowing regions with distinct stellar populations to be modeled separately, reducing the impact of luminous star-forming clumps on the inferred global mass. Overall, these trends confirm that stellar mass bias is closely tied to sSFR, thus SFH, and highlight the importance of spatially resolved measurements for reliably probing the underlying stellar populations. However, the stellar mass bias cannot be fully eliminated with spatially resolved photometry, because even individual Voronoi regions or pixels may still experience outshining, when they contain light from multiple stellar populations.

\subsection{Spatial Mass-to-Light Variations and Stellar Mass Bias}\label{subsec: disc mass-to-light and stellar mass bias} 

Extending previous work, we explore the spatial structure of the outshining effect in high-sSFR galaxies. Galaxies exhibiting significant stellar mass bias also show a broader distribution of resolved mass-to-light ($M_{\ast}/L_V$) ratios, with the largest stellar mass underestimates occurring in systems with the greatest standard deviation in $\log_{\mathrm{10}}(M_{\ast}/L_V)$. This increased dispersion reflects stronger spatial variations in the underlying stellar populations, arising from the coexistence of young, luminous star-forming regions and older, fainter stellar components. 

Consistent with the trends presented in \Cref{fig: bias stellar mass vs std mass-to-light ratios}, the correlation between stellar mass bias and the standard deviation of $\log_{\mathrm{10}}(M_{\ast}/L_V)$ is strongest for galaxies with the highest sSFR ($-7 < \log_{\mathrm{10}}(\mathrm{sSFR}) < -8$)- having a Spearman coefficient of $\rho_{\rm S} = -0.37$. 
The correlation weakens substantially in the next sSFR bin ($-8 < \log_{\mathrm{10}}(\mathrm{sSFR}) < -9$) with $\rho_{\rm S}=-0.01$ and changes sign in the two lowest sSFR bins ($-9 < \log_{\mathrm{10}}(\mathrm{sSFR}) < -10$ and $-10 < \log_{\mathrm{10}}(\mathrm{sSFR}) < -14$) with $\rho_{\rm S}=0.35$ in both cases, where larger $M_{\ast}/L_V$ dispersion is instead associated with increasingly positive stellar mass biases. 

While the negative correlation at high sSFR is consistent with the expected outshining of older stellar populations by young, luminous regions, the positive correlation at low sSFR indicates that the impact of spatially heterogeneous stellar populations on integrated mass estimates is not described by outshining alone. Galaxies with greater spatial variation in $M_{\ast}/L_V$ exhibit increasingly positive stellar mass biases at low sSFR. Thus, while spatially heterogeneous stellar populations are associated with biases in integrated stellar mass estimates, the direction of the bias depends on the sSFR regime. 

The presence of spatially heterogeneous stellar populations is consistent with previous studies, including \cite{2025arXiv251202440M}, which reported significant colour gradients in galaxies out to $z \sim 2.5$, indicative of multiple stellar populations. Together, these results demonstrate that spatially resolved photometry is essential for characterizing variations in stellar populations and for reducing the biases inherent in integrated stellar mass estimates.

\subsection{Impact of Medium Bands on Stellar Mass Estimates}\label{subsec: disc impact medium bands}

Medium-band photometry can significantly affect integrated stellar mass measurements by better constraining the continuum and separating emission lines from the continuum. We find that integrated stellar masses are more sensitive to the presence or absence of medium bands, whereas spatially resolved photometry largely mitigates this effect. This subsection discusses these differences and their implications for both SED-fitting and survey design.

Interestingly, the largest differences in stellar mass with and without medium bands (see \Cref{fig: int res ratio medium band}) occur for intermediate sSFR galaxies. In these systems, a mix of old and young stellar populations broadens SED degeneracies, unlike the highest sSFR galaxies dominated by very young stars. This effect highlights how the combination of stellar population complexity and SED-fitting assumptions drives the sensitivity of integrated masses to medium-band coverage (\citeauthor{2024ApJ...967L..17S} \citeyear{2024ApJ...964..177T}).

Therefore, although medium-band filters are essential for constraining integrated stellar masses, spatially resolved photometry largely mitigates these effects, emphasizing the complementary importance of survey design and spatial resolution. Note that this analysis is made only for high S/N galaxies, and so additional work would be needed to see if spatially resolved broadband photometry could mitigate problems with integrated broadband for galaxies with much lower S/N, which are frequently used in unresolved studies.

\subsection{Lensing Magnification and Stellar Mass}\label{subsec: disc impact magnification}
In five out of 10 fields (\textit{CLU}/cluster fields), galaxies are magnified by gravitational lensing. The magnification can vary from almost negligible values to higher magnifications, 582 out of 1232 \textit{CLU}-field galaxies have $\mu > 2$. \Cref{fig: rgb and 2d maps small bias} and \Cref{fig: rgb and 2d maps large bias} demonstrate examples of galaxies with different magnification factors. Galaxy ID 2101179 stands out for having $\mu = 7.68$. After inspection, this galaxy, as well as several others, seems to experience differential magnification. However, in the spatially resolved analyses, we apply an uniform $\mu$ to all spatial regions of magnified galaxies. Mapping to the non-magnified source plane and applying spatially resolved photometry would provide a more accurate representation of the intrinsic galaxy, but is beyond the scope of this paper and represents an important direction for future work.

\section{Conclusions}\label{sec: conclusion}
In this work, we apply \texttt{Dense Basis} to perform both integrated and spatially resolved SED-fitting on $\sim$3000 galaxies at 1 $<$ $z$ $<$ 9 in the CANUCS, Technicolor, and JUMPS fields. Spatially resolved photometry is constructed applying \texttt{Voronoi} on the F444W imaging, using a threshold of SNR$_{\mathrm{bin}}$ = 15 to ensure uniformly high-quality spatial regions. We investigate the outshining effect to identify the signatures of distinct stellar populations and to assess its correlation with star-forming activity.

\Cref{fig: rgb and 2d maps small bias} and \Cref{fig: rgb and 2d maps large bias} present examples of the two-dimensional stellar mass, dust, and star-formation maps for galaxies both affected and not affected by the outshining effect. We further examine how the outshining effect correlates with surface brightness, and galaxy compactness in \Cref{fig: bias sSFR mass size zphot}, and with variations in mass-to-light ratios in \Cref{fig: rgb hist and 2d maps small bias}, \Cref{fig: rgb hist and 2d maps large bias}, and \Cref{fig: bias stellar mass vs std mass-to-light ratios}.

\begin{enumerate}[label=\arabic*., left=0pt]
    \item 
    We find a correlation between stellar mass bias and the sSFR of a galaxy. The systematic bias is most apparent in low-mass systems ($\log_{10}(M_{\ast}/M_{\odot}) < 9.5$): these galaxies generally have a higher sSFR, making them dominated by bright, young stellar populations. This result is consistent with the findings of \cite{2018MNRAS.476.1532S}. However, the median bias is similar when separating the sample by stellar mass, with values of $-0.03$ dex and $0.03$ dex for the low- and high-mass samples, respectively. In contrast, the median bias varies substantially across sSFR bins, decreasing from $-0.384$ dex in the highest sSFR bin to approximately zero at lower sSFRs (\Cref{tab: median bias ssfr bin}). As shown in \Cref{fig: bias sSFR mass size zphot}, the outshining effect therefore does not affect all galaxies uniformly, but is concentrated in specific populations with high sSFR and strong spatial variations in their stellar populations. These results highlight the importance of accounting for spatially heterogeneous stellar populations when interpreting stellar masses derived from integrated photometry.
    
    \item 
    We find that the stellar mass bias decreases for compact, high surface-brightness galaxies (lower panels of \Cref{fig: bias sSFR mass size zphot}). These galaxies may still experience outshining, but it happens on smaller scales than JWST resolution, such that individual pixels or Voronoi bins contain a mixture of stellar populations. Although \texttt{Dense Basis} uses a non-parametric SFH that should help mitigate outshining within compact regions, spatially resolved non-parametric SFHs alone do not fully eliminate the effect. While this work does not test parametric SFHs, \cite{2025MNRAS.542.2998H} showed that varying the SFH parameterisation does not necessarily resolve the outshining effect. 
    
    \item 
    Our large sample confirms that medium-band photometry is essential for accurate stellar mass estimates in integrated photometry, consistent with the findings of \cite{2024ApJ...967L..17S}. \Cref{fig: int res ratio medium band} shows that, across all five CANUCS clusters and their parallel fields, with the exclusion of MACS0417-\textit{CLU} and MACS1423-\textit{CLU}, intermediate- and low sSFR galaxies tend to have overestimated masses when medium-bands are excluded, whereas a smaller subset of high sSFR galaxies exhibit underestimated masses. The standard deviation of the bias distribution- between inclusion and exclusion of medium bands- is $\sigma_{\mathrm{int}} = 0.33$ dex for integrated photometry and $\sigma_{\mathrm{res}} = 0.15$ dex for spatially resolved photometry. Importantly, spatially resolved photometry mitigates these offsets, reducing the dependence on medium-band filters for robust stellar mass estimates.
    
    \item 
    We examine the distribution of the mass-to-light ratio ($M_{\ast}/L_{V}$), which quantifies the stellar mass relative to the emitted light and provides a direct diagnostic of the outshining effect. Young stellar populations produce bright regions with low $M_{\ast}/L_{V}$, whereas older stellar populations are fainter and characterised by higher $M_{\ast}/L_{V}$. As shown in \Cref{fig: rgb hist and 2d maps small bias} and \Cref{fig: rgb hist and 2d maps large bias}, galaxies strongly affected by outshining exhibit a broader distribution of $M_{\ast}/L_{V}$, reflecting the coexistence of these distinct stellar populations. Consequently, these systems display a larger standard deviation in $\log(M_{\ast}/L_{V})$, indicative of stronger spatial variations in the underlying stellar populations. The top panels of \Cref{fig: bias stellar mass vs std mass-to-light ratios} further show that galaxies with the highest sSFR ($-7 < \log_{\mathrm{10}}(\mathrm{sSFR}) < -8$) exhibit the largest dispersion in $\log_{\mathrm{10}}(M_{\ast}/L_{V})$ and the greatest stellar mass bias, while this trend becomes progressively weaker at lower sSFR. This behaviour is consistent with the outshining effect being strongest in actively star-forming galaxies, where young, luminous regions dominate the integrated light and bias stellar mass estimates towards lower values.

\end{enumerate}

Our results provide a large sample of resolved stellar mass estimates and two-dimensional maps of key galaxy parameters, illustrating the effectiveness of resolved photometry with JWST. Previous studies have shown that stellar mass bias increases with star-formation activity, particularly in low-mass, star-forming galaxies, and that medium-band filters are essential in integrated photometry for accurately constraining the continuum. Using a substantially larger sample than previous work, we confirm these trends and further show that galaxies affected by outshining exhibit pronounced variations in their $M_{\ast}/L_V$, corresponding to larger stellar mass biases. In addition, we find that compact galaxies do not exhibit a measurable bias, as individual pixels or Voronoi regions in these systems contain a wide mixture of stellar populations; consequently, spatially resolved photometry cannot mitigate outshining in these cases. Furthermore, we find that spatially resolved photometry effectively mitigates the biases in stellar-mass estimates that arise when medium-band filters are not available. Overall, resolved photometry proves essential for uncovering internal stellar population variations and for reducing systematic biases in integrated mass measurements.

\section*{Acknowledgements} 
This research was supported by grants 23JWGO2A13 and 24JWGO3A04 from the Canadian Space Agency (CSA), and funding from the Natural Sciences and Engineering Research Council of Canada (NSERC). AM acknowledges support from the Yavin Family Fund. YA acknowledges support from the Dunlap Institute, funded through an endowment established by the David Dunlap family and the University of Toronto. MB acknowledges support from the Slovenian national research agency ARRS through grant N1-0238 and the program HST-GO-16667, provided through a grant from the STScI under NASA contract NAS5-26555. SCB acknowledges that support for this work was provided by the National Science Foundation Astronomy \& Astrophysics
Postdoctoral Fellowship Award No. 2503156. DM acknowledges generous support from the Leonard and Jane Holmes Bernstein Professorship in Evolutionary Science. Support for programs JWST-GO-03362 and JWST-GO-05890, provided through a grant from the STScI under NASA contract NAS5-03127, is acknowledged. This research used the Canadian Advanced Network For Astronomy Research (CANFAR) operated in partnership by the Canadian Astronomy Data Centre and The Digital Research Alliance of Canada with support from the National Research Council of Canada the Canadian Space Agency, CANARIE and the Canadian Foundation for Innovation. I also want to thank Rachel Bezanson for helpful discussions on the stellar surface mass density.\\
\textit{Facilities}: \textit{JWST} (NASA/ESA/CSA), \textit{HST} (NASA/ESA). \\
\textit{Software}: \texttt{ASTROPY} (\citeauthor{2018AJ....156..123A} \citeyear{2018AJ....156..123A}), \texttt{MATPLOTLIB} (\citeauthor{4160265} \citeyear{4160265}), \texttt{NUMPY} (\citeauthor{2011CSE....13b..22V} \citeyear{2011CSE....13b..22V}), \texttt{CMASHER} (\citeauthor{2020JOSS....5.2004V} \citeyear{2020JOSS....5.2004V}).


\appendix
\twocolumngrid

\section{Voronoi Binning \& Signal-to-Noise Ratio Thresholds}\label{sec: appendix voronoi and snr threshold}

Spatially resolved measurements are obtained via Voronoi binning of pixel-level fluxes. We adopt a target signal-to-noise ratio per bin of $S/N_{\rm bin} = 15$, which provides a balance between reliable specific star formation rate estimates and sufficient spatial resolution. Stellar masses can be recovered with reasonable accuracy even at lower thresholds ($S/N_{\rm bin} \simeq 5$), whereas sSFRs require higher-quality bins ($S/N_{\rm bin} \simeq 20$) to limit uncertainties.

\begin{figure}[h]
    \centering
    \includegraphics[width=\columnwidth]{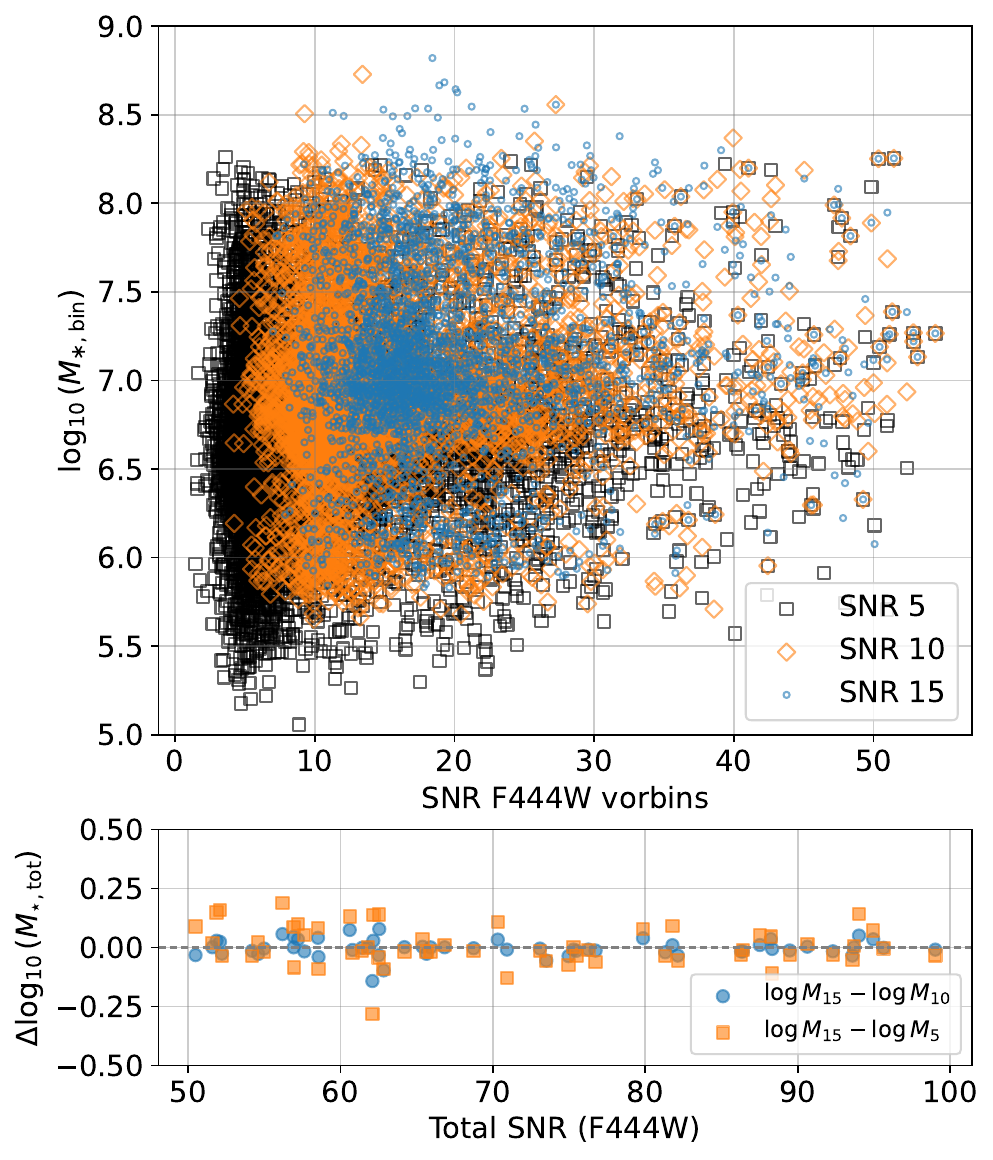}
    \hfill
    \caption{Different colours and symbols indicate the resolved SNR binning used in the stellar-mass reconstruction. Upper panel: resolved stellar-mass bins (log $m_{50}$) as a function of resolved SNR for a subset of galaxies that are common to all two resolved SNR binning choices (SNR = 5, 10, 15) and that satisfy the low-SNR selection in F444W $<$ 100. Lower panel: total stellar mass, obtained by summing the resolved stellar-mass bins for each galaxy, plotted as a function of the integrated F444W signal-to-noise ratio measured from aperture photometry according to DR1. The total stellar masses show some scatter between the different resolved SNR binning choices, with the differences being largest between SNR = 5 and 15, but remain within $\sim$0.25 dex. This indicates that the adopted SNR threshold has a relatively small effect on the integrated stellar mass compared to the scatter between individual binning choices. We therefore adopt SNR = 15, which provides improved SED-fitting accuracy while preserving consistent integrated stellar-mass estimates compared to the lower SNR thresholds.}
    \label{fig: log stelm vs snr of vorbins}
\end{figure}

To test the effect of the SNR threshold on resolved stellar mass estimates, we generated tessellations with $S/N_{\rm bin} = 5$, 10, and 15. The resulting total stellar masses are consistent across the three thresholds, with differences of $\lesssim 0.25$ dex and no clear systematic offset. The larger scatter at lower SNR is consistent with the less constrained SED fits in lower-SNR regions. Since we consider differences within $\sim0.3$ dex to be consistent with random scatter, the choice of $S/N_{\rm bin} = 15$ does not introduce a significant systematic difference in the total resolved stellar mass or the derived stellar mass bias. We therefore adopt $S/N_{\rm bin} = 15$ as a conservative threshold for the resolved analysis, while noting that the choice of threshold affects the spatial resolution of low-surface-brightness features and localized outshining. A visual comparison of tessellations for the different thresholds is presented in \Cref{fig: log stelm vs snr of vorbins}.

\section{Flowchart Method: Integrated vs Resolved Photometry}\label{sec: appendix flowchart method}

\begin{figure}[h]
    \centering
    \textbf{Integrated Photometry\quad|\quad Resolved Photometry}\\[.1em]
    \includegraphics[width=\columnwidth]{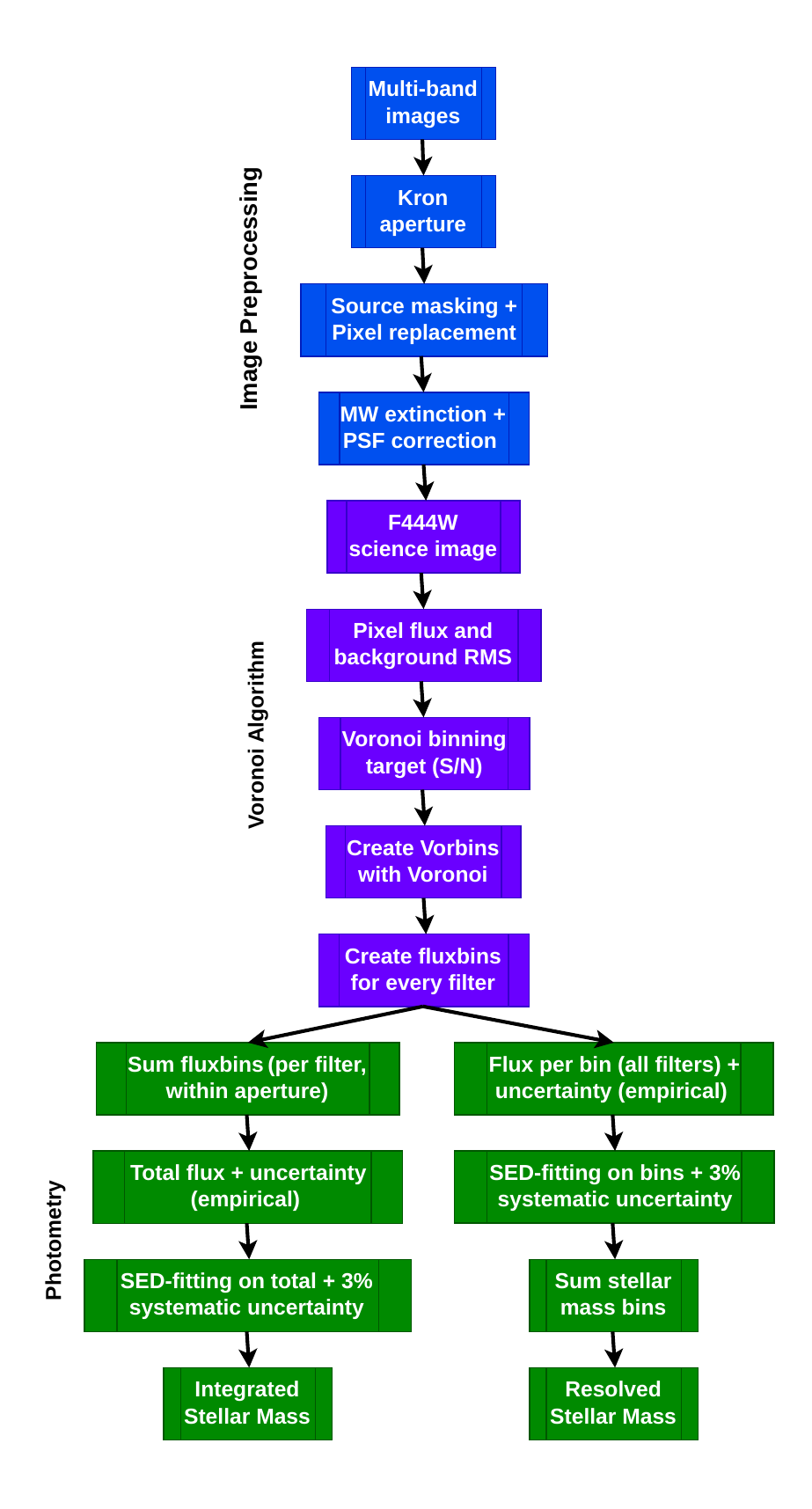}
    \hfill
    \caption{Flowchart of the methodology used to derive integrated and resolved stellar masses (see Section \ref{sec: appendix flowchart method} for details). After common preprocessing, Voronoi binning is performed on the F444W image to define fluxbins that are applied across all filters. The pipeline then splits into two approaches: integrated photometry, where fluxes are summed prior to SED-fitting, and resolved photometry, where SED-fitting is performed per bin and stellar masses are summed. Photometric uncertainties are derived empirically from empty aperture measurements and applied consistently in both cases.}
    \label{fig: flowchart method}
\end{figure}

\Cref{fig: flowchart method} summarizes the workflow used to derive integrated and resolved stellar masses. All images are first processed through a common preprocessing stage, including source masking, pixel replacement, aperture definition, and corrections for Milky Way extinction and PSF effects applied consistently across all filters.

Voronoi binning is performed on the F444W image to define the spatial segmentation. The binning uses a background pixel RMS estimate to set the target signal-to-noise ratio, producing a set of Voronoi bins (hereafter "fluxbins"). This segmentation is then applied uniformly to all filters to extract photometric measurements.
\\
From these fluxbins, two approaches are followed. For integrated photometry, fluxes are summed across all bins within the aperture in each filter to obtain total fluxes. For resolved photometry, fluxes are retained at the bin level. In both cases, photometric uncertainties are derived using an empirical method based on empty aperture measurements, accounting for correlated noise and scaled to the effective areas of the bins or aperture, with an additional 3\% systematic uncertainty included.
\\
Stellar masses are derived using Dense Basis SED-fitting. In the integrated case, SED-fitting is performed once on the total fluxes. In the resolved case, SED-fitting is performed independently for each bin, and the resulting stellar masses are summed. This framework ensures that both approaches are treated consistently, differing only in the order of flux summation and SED-fitting.

\section{Clump Fraction}\label{sec: appendix clump fraction}
\begin{figure}[h!]
    \centering
    \includegraphics[width=\columnwidth]{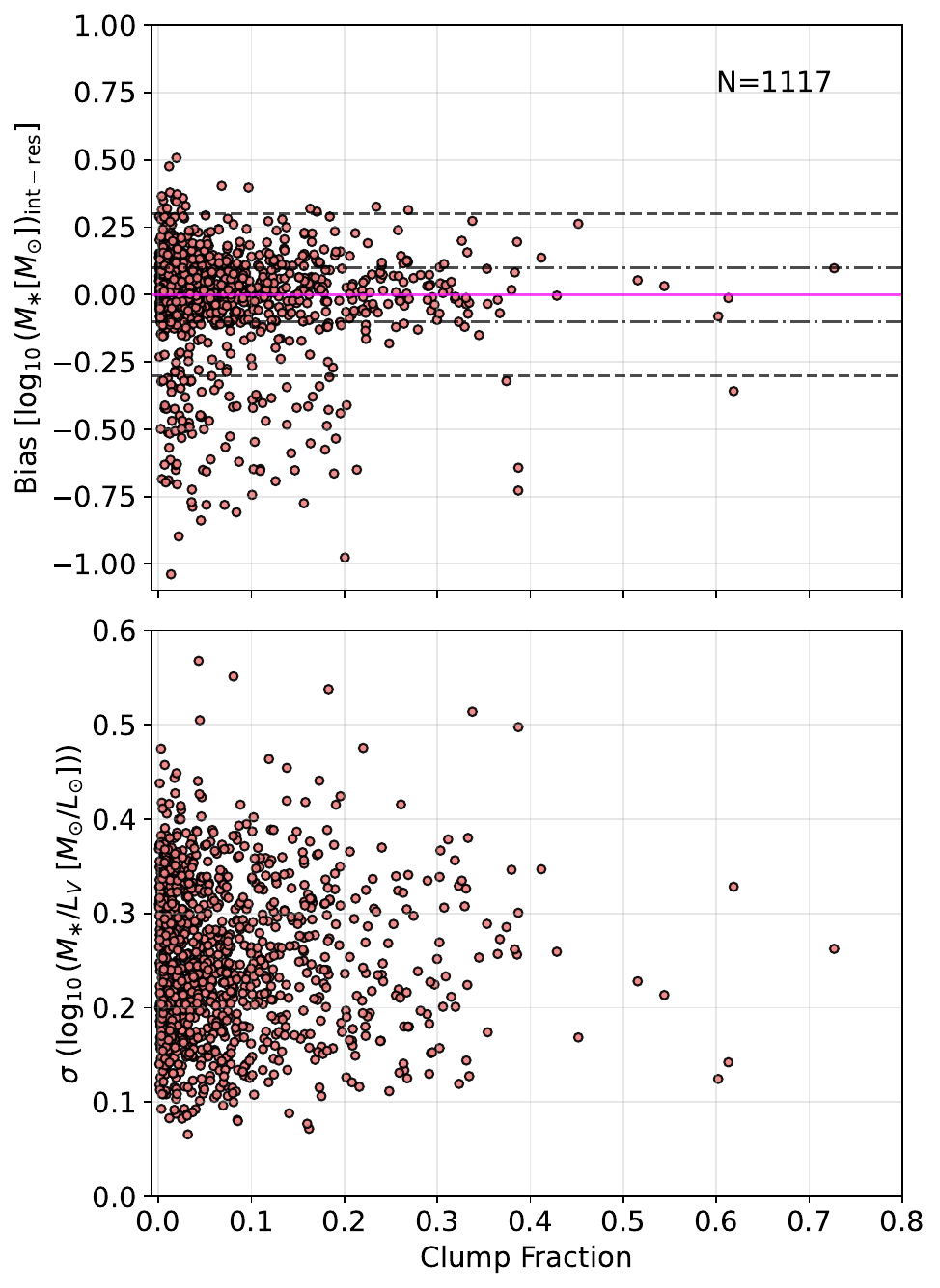}
    \hfill
    \caption{Stellar mass bias, defined as $\log M_{\mathrm{int}} - \log M_{\mathrm{res}}$ (top panel), and the spread of the spatially resolved $\log(M_\ast/L_V)$ distribution (bottom panel), shown as a function of the stellar-mass clump fraction. The clump fraction is derived from clump identification in rest-frame $U$-band maps and is defined as the fraction of the total stellar mass associated with clump regions relative to the integrated stellar mass of the galaxy. Horizontal dashed lines in the top panel mark zero bias and reference levels at $\pm 0.1$ and $\pm 0.3$ dex. No strong correlation is observed in either panel, indicating that the presence of clumpy structure alone does not necessarily result in significant stellar mass bias or high $\sigma$ in the mass-to-light ratio.
    }
    \label{fig: bias stelm clump fraction}
\end{figure}

We investigate whether the presence of morphological substructure contributes to biases in integrated stellar mass estimates by examining the relation between stellar mass bias and the stellar-mass clump fraction. The clump fraction is derived from clump identification in rest-frame $U$-band images and is defined as the fraction of the total stellar mass associated with clump regions relative to the integrated stellar mass of the galaxy.
\\
Figure \ref{fig: bias stelm clump fraction} shows stellar mass bias as a function of clump fraction. No strong correlation is observed: galaxies with higher clump fractions do not systematically exhibit larger differences between integrated and spatially resolved stellar masses. This indicates that the presence of clumpy structure alone is not sufficient to produce significant stellar mass bias.
\\
This result is consistent with the interpretation that stellar mass bias is primarily driven by internal variations in the mass-to-light ratio rather than by morphology alone. While clumpy galaxies may host young, luminous star-forming regions, significant bias arises only when these coexist with spatially segregated stellar populations spanning a broad range of $M_\ast/L$. Spatially resolved mass-to-light ratio diagnostics therefore provide a more direct tracer of stellar mass bias than clump fraction.
\FloatBarrier
\newpage

\bibliography{references}
\bibliographystyle{aasjournalv7.1}



\end{document}